\documentclass[apj]{emulateapj}
\usepackage{natbib}
\usepackage{multirow}
\usepackage{graphicx}
\usepackage[pdftex,                %%% hyper-references for pdflatex
bookmarks=true,                    %%% generate bookmarks ...
bookmarksnumbered=true,            %%% ... with numbers
colorlinks=true,            % links are colored
citecolor=blue,         % green   % color of cite links
linkcolor=blue,     %cyan,         % color of hyperref links
menucolor=blue,         % color of Acrobat Reader menu buttons
urlcolor=cyan,       % color of page of \url{...}
linkbordercolor={0 0 1},           %%% blue frames around links
pdfborder= {0 0 1},
frenchlinks=True]{hyperref}

\newcommand{\fig}[5]{
        \begin{figure}[!bt]
        \begin{center}
        \includegraphics[#3]{#2}
      \end{center}
      
        \renewcommand{\baselinestretch}{1}
        \vspace*{-.3in}
        \caption[#4]{#5}
        \label{fig:#1}
        \end{figure}}

\newcommand{\figtwocol}[5]{
        \begin{figure*}[!bt]
        \begin{center}
        \includegraphics[#3]{#2}
      \end{center}
      \renewcommand{\baselinestretch}{1}
        \vspace*{-.3in}
        \caption[#4]{\footnotesize #5}
        \label{fig:#1}
        \end{figure*}}

\newcommand{\uab}{\ensuremath{\upsilon}~And~b}

\newcommand{\uandb}{upsilon~Andromeda~b}
\newcommand{\hdtwo}{HD~209458}
\newcommand{\hdtwob}{HD~209458b}
\newcommand{\hdone}{HD~189733}
\newcommand{\hdoneb}{HD~189733b}

\newcommand{\rjup}{\ensuremath{R_{J}}}
\newcommand{\mjup}{\ensuremath{M_{J}}}

\newcommand{\jeclipses}{\ensuremath{0.338\% \pm 0.026\%}}

\newcommand{\jtransits}{\ensuremath{1.484\% \pm 0.033\%}}

\newcommand{\systemfluxes}{\ensuremath{18.7\pm 0.4}~mJy}
\newcommand{\fixedperiod}{\ensuremath{3.52474550 \pm 0.00000018~\textrm{d}}}

\newcommand{\fittransitcov}{\ensuremath{-5.652 \times 10^{-9} \textrm{~d}^2}}
\newcommand{\fittransitephemeris}{\ensuremath{2453549.2075 \pm 0.0013 \textrm{~d}}}
\newcommand{\fittransitperiod}{\ensuremath{3.5247537 \pm 0.0000049 \textrm{~d}}}
\newcommand{\fittransitdifference}{\ensuremath{8.2 \times 10^{-6} \textrm{~d}}}
\newcommand{\fittransitdifferences}{0.71~s}
\newcommand{\deltatc}{\ensuremath{32 \pm 129 \textrm{~s}}} % earlier than expected (accounts for 47 s travel time)
\newcommand{\timediff}{\ensuremath{0.79 \pm 0.94 \textrm{~s}}} % P_transit - P_eclipse
\newcommand{\fiteclipseperiod}{\ensuremath{3.5247445 \pm 0.0000097 \textrm{~d}}}
\newcommand{\fiteclipsedifference}{\ensuremath{0.99 \times 10^{-6}\textrm{~d}}}
\newcommand{\fiteclipsedifferences}{\ensuremath{0.086\textrm{~s}}}
\newcommand{\ecosomega}{\ensuremath{0.00004 \pm 0.00033}}
\newcommand{\daysidetemps}{\ensuremath{1320\textrm{~K} \pm 80\textrm{~K}}} % 2011-11-04

\newcommand{\cten}{C10}     % \cite{crossfield:2010}
\newcommand{\ctenp}{(C10)}  % \citep{crossfield:2010}
\newcommand{\water}{\ensuremath{\textrm{H}_2\textrm{O}}}

\providecommand{\eprint}[1]{\href{http://arxiv.org/abs/#1}{#1}}
\providecommand{\adsurl}[1]{\href{#1}{ADS}}
\providecommand{\emailh}[1]{\protect{\href{mailto:#1}{#1}}}
\def\aap{{A\&A}}		% Astronomy and Astrophysics
\def\apj{{ApJ}}			% Astrophysical Journal
\def\apjl{{ApJ}}		% Astrophysical Journal, Letters
\def\apjs{{ApJS}}		% Astrophysical Journal, Supplement
\def\pasp{{PASP}}		% Publications of the ASP

\def\mnras{{MNRAS}}
\def\nat{{Nature}}

\shortauthors{Crossfield et al.}
\shorttitle{24\,\micron\ Observations of HD~209458b}
\begin{document}

\title{Spitzer/MIPS 24\,\micron\ Observations of HD~209458b: 3
  Eclipses, 2.5 Transits, and a Phase Curve Corrupted by Instrumental
  Sensitivity Variations. }

\slugcomment{ApJ accepted: 2012 April 11}

\author{
Ian J. M. Crossfield\altaffilmark{1},
Heather Knutson\altaffilmark{2}, 
Jonathan Fortney\altaffilmark{3}$^,$\altaffilmark{4},
Adam P. Showman\altaffilmark{5},
Nicolas B. Cowan\altaffilmark{6}$^,$\altaffilmark{7},
Drake Deming\altaffilmark{8}}

\altaffiltext{1}{Department of Physics \& Astronomy, University of
  California, Los Angeles, CA 90095, USA \emailh{ianc@astro.ucla.edu}}

\altaffiltext{2}{Caltech Division of Geological and Planetary
  Sciences, Pasadena, CA 91125, USA}

\altaffiltext{3}{Department of Astronomy and Astrophysics, University
  of California, Santa Cruz, CA 95064, USA}

\altaffiltext{4}{Alfred P. Sloan Research Fellow}

\altaffiltext{5}{Department of Planetary Sciences and Lunar and
  Planetary Laboratory, The University of Arizona, Tucson, AZ 85721,
  USA}

\altaffiltext{6}{Center for Interdisciplinary Exploration and Research
  in Astrophysics (CIERA) and Department of Physics \& Astronomy,
  Northwestern University, Evanston, IL 60208, USA}

\altaffiltext{7}{CIERA Postdoctoral Fellow}

\altaffiltext{8}{Department of Astronomy, University of Maryland,
  College Park, MD 20742, USA}

\begin{abstract}
  We report the results of an analysis of all Spitzer/MIPS
  24\,\micron\ observations of \hdtwob, one of the touchstone objects
  in the study of irradiated giant planet atmospheres. Altogether we
  analyze two and a half transits, three eclipses, and a 58-hour
  near-continuous observation designed to detect the planet's thermal
  phase curve.  The results of our analysis are: (1) A mean transit
  depth of \jtransits, consistent with previous measurements and
  showing no evidence of variability in transit depth at the 3\%\
  level. (2) A mean eclipse depth of \jeclipses, somewhat higher than
  that previously reported for this system; this new value brings
  observations into better agreement with models.  From this eclipse
  depth we estimate an average dayside brightness temperature of
  \daysidetemps; the dayside flux shows no evidence of variability at
  the 12\%\ level. (3) Eclipses in the system occur \deltatc\ earlier
  than would be expected from a circular orbit, which constrains the
  orbital quantity $e \cos \omega $ to be \ecosomega. This result is
  fully consistent with a circular orbit and sets an upper limit of
  140~m~s$^{-1}$ ($3\sigma$) on any eccentricity-induced velocity
  offset during transit. The phase curve observations
  (including one of the transits) exhibits an anomalous trend similar
  to the detector ramp seen in previous Spitzer/IRAC observations; by
  modeling this ramp we recover the system parameters for this
  transit.  The long-duration photometry which follows the ramp and
  transit exhibits a gradual $ \sim0.2\% $ decrease in flux over $
  \sim 30$~hr. This effect is similar to that seen in pre-launch
  calibration data taken with the 24\,\micron\ array and is better fit
  by an instrumental model than a model invoking planetary emission.
  The large uncertainties associated with this poorly-understood,
  likely instrumental effect prevent us from usefully constraining the
  planet's thermal phase curve.  Our observations highlight the need
  for a thorough understanding of detector-related instrumental
  effects on long time scales when making the high-precision
  mid-infrared measurements planned for future missions such as EChO,
  SPICA, and JWST.

\end{abstract}


\section{Introduction}
Most known extrasolar planets were discovered via the radial velocity
technique -- in which the Doppler wobble of a star indicates an
orbiting planet -- and/or by the transit method -- in which periodic
dimming of a star indicates a planet that crosses in front of the
stellar disk.  Owing to the observational biases of these techniques,
the first planets thus discovered were the large, massive objects on
few-day orbits commonly known as hot Jupiters
\citep{mayor:1995,henry:2000,charbonneau:2000}.  Their large sizes and
high temperatures make these objects excellent candidates for the
study of their dayside emission when the planet is occulted by the
star \citep{deming:2005a,charbonneau:2005}, of their
longitudinally-averaged global emission
\citep{harrington:2006,cowan:2007,knutson:2007b}, and of their atmospheric
opacity via the wavelength-dependent flux diminution during transit
\citep{seager:2000,charbonneau:2002}. These observations have led to
measurements of atmospheric abundances of key molecular species
\citep{madhusudhan:2011}, possible non-equilibrium chemistry
\citep{stevenson:2010}, high-altitude hazes \citep{sing:2009haze}, and
atmospheric circulation \citep{cowan:2011}.

Any discussion of hot Jupiter atmospheres must necessarily mention two
systems in particular.  One, \hdone, is the brightest star known to
host a hot Jupiter \citep{bouchy:2005}.  The other is \hdtwo, the
first known transiting planet \citep{charbonneau:2000,henry:2000} and
the focus of this study.  These are the two touchstone objects in the
study of irradiated giant exoplanets, both because they were
discovered relatively early on and because they orbit especially
bright (as seen from Earth) host stars.  This last point in particular
allows for especially precise characterization of these planets'
atmospheres and permits observations which would provide unacceptably
low signal to noise ratios for fainter systems.

\subsection{The  HD~209458 system}
The star \hdtwo\ is an F8 star roughly 15\,\%\ more massive than the Sun
\citep{mazeh:2000,brown:2001,baines:2008}, with an equivalent
metallicity and slightly higher temperature \citep{schuler:2011}.
It is orbited by \hdtwob, a roughly 1.4\,\rjup, 0.7\,\mjup\ planet in
a 3.5-day, near-circular orbit \citep{southworth:2008,torres:2008}.
The planet's parameters have been substantially improved upon since
its initial discovery
\citep{charbonneau:2000,henry:2000,mazeh:2000}. Two sets of more
recent values \citep{torres:2008,southworth:2008} do not differ
significantly, and we use the former's system parameters in our
analysis when not making our own measurements.

Infrared photometry during eclipses of \hdtwob\ measured from the
ground \citep{richardson:2003b} and with Spitzer \citep{deming:2005a,
  knutson:2008} determines the planet's intrinsic emission spectrum,
and is best fit by atmospheric models in which the planet's
atmospheric temperature increases above $\sim0.1-1$~bar
\citep{burrows:2007,burrows:2008,fortney:2008,madhusudhan:2010}.  Such
temperature inversions are common on hot Jupiters, and a popular
explanation requires the presence of a high-altitude absorber
\citep[e.g.,][]{fortney:2008,burrows:2008}.  The nature of any such
absorber is currently unknown and the subject remains a topic of
active research \citep{desert:2008,spiegel:2009,knutson:2010, madhusudhan:2011b}.

If present, a high-altitude optical absorber is expected to absorb the
incident stellar flux high in the atmosphere where radiative
timescales are short and advection is inefficient
\citep{cowan:2011circ}. Consequently, such planets are expected to
exhibit large day/night temperature contrasts and low global energy
redistribution despite circulation models' ubiquitous predictions of
large-scale superrotating jets on these planets \citep{showman:2002,
  cooper:2005,cho:2008,rauscher:2008,showman:2009,dobbs-dixon:2010,burrows:2010,
  rauscher:2010, thrastarson:2010,heng:2011a,heng:2011b}.
Spitzer/IRAC observations of \hdtwob\ at 8\,\micron\ place an upper
limit on the planet's thermal phase variation of 0.0022
\citep[$3\sigma$;][]{cowan:2007}.  Given the planet's demonstrably low
albedo \citep{rowe:2008} this limit is substantially lower expected if
the planet has a low recirculation efficiency.  In hot Jupiter
atmospheres the dominant 24\,\micron\ molecular opacity source is
expected to be \water, but there is some tension between models and
past observations at this wavelength
\citep[cf.][]{madhusudhan:2010}. Thus our understanding of these
planets' atmospheres remains incomplete.

Recent spectroscopic observations of \hdtwob\ during transit show a
hint of a systematic velocity offset ($2 \pm 1 \textrm{km~s}^{-1}$) of
planetary CO lines during planetary transit \citep{snellen:2010}.  If
confirmed, this offset would be diagnostic of high-altitude winds
averaged over the planet's day/night terminator, and similar
measurements at higher precision could one day hope to spatially
resolve terminator circulation patterns and constrain atmospheric drag
properties \citep{rauscher:2012}.  However, small orbital
eccentricities (specifically, nonzero $e \cos \omega$, where $\omega$
is the longitude of periastron) can also induce a velocity offset in a
planetary transmission spectrum \citep{montalto:2011}.  It is thus
convenient that precise timing of planetary transits and eclipses
directly constrains $e \cos \omega$ \citep[][chapter by
J.~Winn]{seager:2011}.  This provides a further motivation for our
work: to more tightly constrain \hdtwob's orbit via a homogeneous
analysis of a single, comprehensive data set.

In this paper we analyze the full complement of data for the \hdtwo\
system taken with the MIPS 24\,\micron\ camera \citep[which we
hereafter refer to simply as MIPS;][]{rieke:2004} on the Spitzer Space
Telescope.  MIPS has taken previous 24\,\micron\ observations of
exoplanetary transits \citep{richardson:2006,knutson:2009b}, eclipses
\citep{deming:2005a,charbonneau:2008,knutson:2008,knutson:2009a,stevenson:2010},
and thermal phase curves
\citep{harrington:2006,knutson:2009a,crossfield:2010}.  MIPS
operations depended on cryogenic temperatures; since Spitzer's
complement of cryogen has been exhausted there may be no further
exoplanet measurements at wavelengths $>10$\,\micron\ until the
eventual launch of missions such as EChO, SPICA, or the James Webb
Space Telescope (JWST). Our work here describes some of the last
unpublished 24\,\micron\ exoplanet observations, and a further
motivation for our work is to inform the calibration, reduction, and
observational methodologies of future missions' mid-infrared (MIR)
observations.

\subsection{Outline}
This report is organized as follows: in Section~\ref{sec:obs} we
describe the MIPS observations and our approach to measuring precise
system photometry.  In Section~\ref{sec:inst} we describe our efforts
to understand the origin of instrumental sensitivity variations
apparent in the long-duration phase curve observations; these effects
ultimately prevent any measurement of \hdtwob's thermal phase curve.
However, we are able to recover the parameters of the observed
transits and eclipses, and we present these results in
Sec.~\ref{sec:transit} and~\ref{sec:eclipse}, respectively.  Combining
the results of these two analyses allows us to constrain the planet's
orbit (i.e., $e \cos \omega$), and we discuss the implications of
this, and of the total system flux, in Section~\ref{sec:misc}.  We
summarize our conclusions and present some thoughts for future
high-precision MIR observations in Section~\ref{sec:conclusion}.

\section{Observations and Analysis}
\label{sec:obs}

\subsection{Observations}
We reanalyzed all observations of the \hdtwo\ system taken with
Spitzer's MIPS 24\,\micron\ channel:  analysis of one transit, two
eclipses, and the long-duration phase curve observations have remained
unpublished until now. Altogether, we used the data from Spitzer
Program IDs 3405 \citep[PI Seager; published in][]{deming:2005a},
20605 \citep[PI Harrington; published in][]{richardson:2006}, and
40280 (PI Knutson).  Table~\ref{tab:obs} lists the observatory
parameters used for each set of observations.  Collectively these data
comprise 2.5~transits, three eclipses, and a 58-hour set of
near-continuous observations designed to detect the planet's thermal
phase curve.

\subsection{Data Reduction}
Unless stated otherwise we use the same methodology to reduce our data
as described in \cite[][hereafter C10]{crossfield:2010}, performing
PSF-fitting photometry using a 100$\times$ super-sampled MIPS
PSF\footnote{Generated using Tiny Tim; available at
  \url{http://ssc.spitzer.caltech.edu/}} modeled using a 6070~K
blackbody spectrum simulated at the center of the MIPS field of view.
We vary the size of the synthetic aperture used to calculate our
PSF-fitting photometry, and find that a square, $21 \times 21$~pixel
aperture minimizes photometric variations.  During MIPS observations
the target star is dithered between fourteen positions on the detector
\citep[][Section~8.2.1.2]{SSC:2007}, and we fit the data from all
dither positions simultaneously as described below.

As noted by \cten, the MIPS 24\,\micron\ detector appears to suffer
from low-amplitude temporal variations in the diffuse background,
presumably owing to small amounts of scattered light in the
instrument.  Because this could affect the flat-fielding performed by
the MIPS reduction pipeline, we create an empirical flat field by
taking a pixel-by-pixel median of all the individual frames after
masking the region containing the target star. After constructing this
flat field we extract photometry (a) after subtracting the master flat
field from each frame, and (b) after dividing each frame by the
normalized-to-unity master flat field.  Both of these give photometry
that is very slightly less noisy (RMS reduced by $\lesssim 1$\%) than
photometry that does not use an additional flat field correction.
Subtracting by the empirical flat-field prior to computing PSF-fitting
photometry results in a lower residual RMS and so we use this approach
for all our data; ultimately our choice of flat field does not change
our final results.

We extract the heliocentric Julian Date (HJD) from the timing tags in
each BCD data file, and then convert the HJD values to BJD$_{TDB}$
using the IDL routine \texttt{hjd2bjd}\footnote{\protect{Available at
    \url{http://astroutils.astronomy.ohio-state.edu/time/}}}
\citep{eastman:2010}.  These new time stamps have an estimated accuracy
of one second \citep{eastman:2010}, which is small compared to our
final ephemeris uncertainties of roughly one and four minutes for
transits and eclipses, respectively.

\subsection{Approach to Model Fitting}
The MIPS dither pattern introduces systematic offsets of $\lesssim
1\%$ \citep{deming:2005a} in the photometry at each dither position.
We follow the methodology of \cten\ and explicitly fit for this effect
by multiplying the modelled flux for each visit at dither position $i$
by the factor $(1 + c_i)$. We further impose the constraint that these
corrections do not change the absolute flux level, and so define $c_0$
such that the quantity $\prod_i \left(1 + c_{i} \right) $ is equal to
unity.  We ultimately find that the $c_i$ are similar, but not
constant, from one epoch to the next.

In all cases we determine best-fit model parameters using the Python
simplex minimization routine
\texttt{scipy.optimize.fmin}\footnote{\protect{Available at
    \url{http://scipy.org/}}}.  We assess parameter uncertainties
using a Markov Chain Monte Carlo implementation of the
Metropolis-Hastings algorithm
(\texttt{analysis.generic\_mcmc}\footnote{\protect{Currently available
    at \url{http://www.astro.ucla.edu/~ianc/python/}}}),
then take as uncertainties the range of values (centered on the
best-fit value) that enclose 68.3\%\ of the posterior distribution. We
verify by eye that the Markov chains are well-mixed; the resulting
one-dimensional posterior distributions are unimodal, symmetric, and
approximately Gaussian unless stated otherwise.

\section{Calibration and Instrument Stability}
\label{sec:inst}
\subsection{The Ramp}
Before we present the results of our model fits, we discuss two
photometric variations that we conclude to be of instrumental origin.
The \hdtwo\ system flux measured from our 2008 observations, shown in
Figure~\ref{fig:crossfield_raw_timeseries}, exhibits a steep rise
during the first 10-12 hours in which the measured system flux
increases by $\sim2\%$. This ramp appears similar to that seen in
photometric observations taken with Spitzer/IRAC and Spitzer/IRS
\citep{charbonneau:2005, deming:2006, knutson:2007b}. The IRAC ramp is
the better studied, and is thought to result from charge-trapping in
the detector \citep[cf.][]{knutson:2007b,agol:2010}.  According to this explanation,
a substantial fraction of photoelectrons liberated early in the
observations become trapped by detector impurities, resulting in a
lower effective gain for the detector. Eventually all charge-trapping
sites become populated and the detector response asymptotes to a
constant level. As the IRAC~8\,\micron, IRS~16\,\micron, and MIPS
detectors are all constructed of Si:As it is conceivable that the MIPS
ramp we observe has a similar origin in charge-trapping.

\fig{crossfield_raw_timeseries}{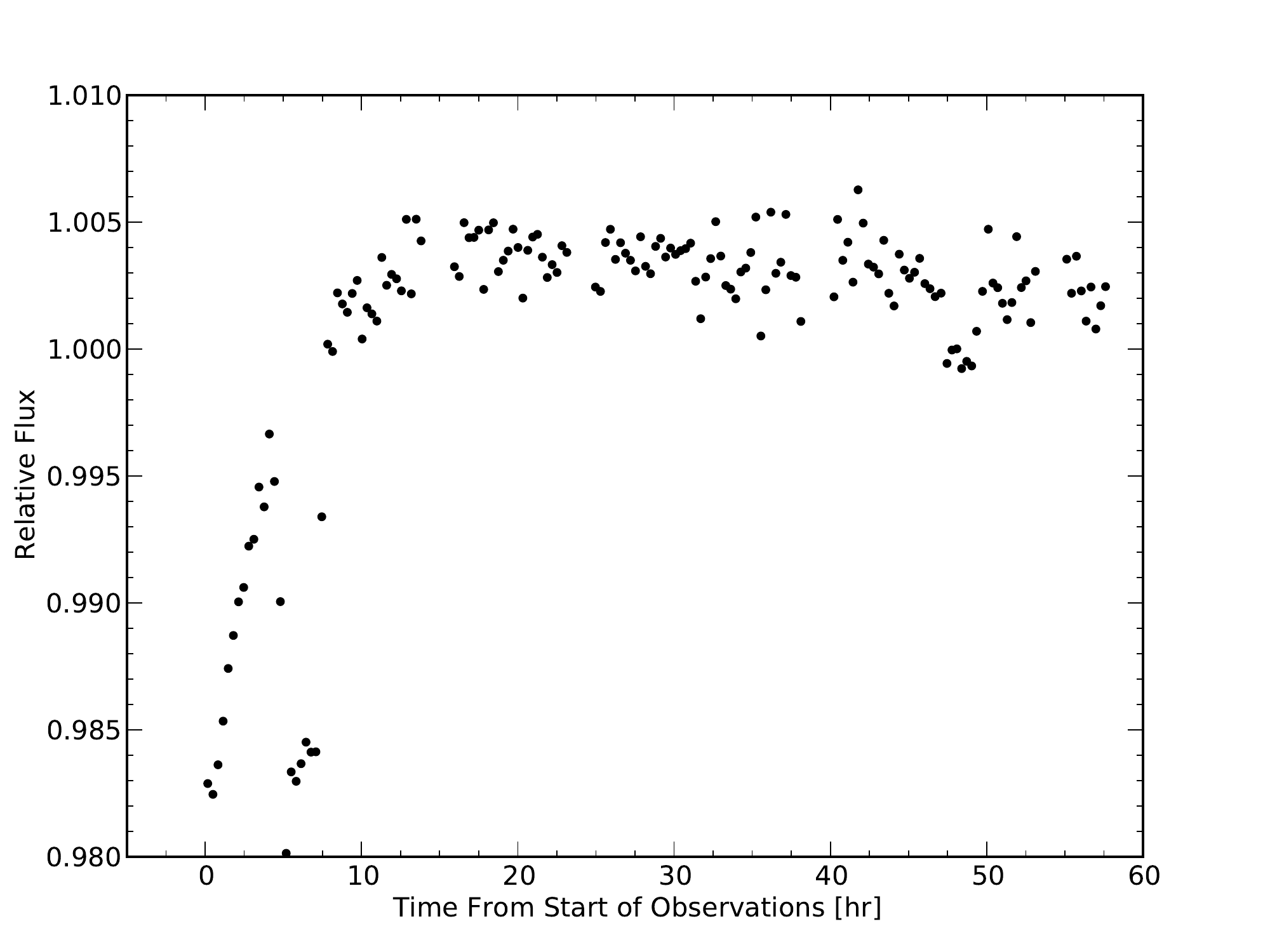}{width=3in}{}{MIPS
  24~\micron\ photometry of the \hdtwo\ system, showing the detector
  ramp (0--10~h), transit (5~h), and eclipse (48~h). For plotting
  purposes the data have been binned to lower temporal resolution.  A
  slight ($\sim0.2\%$) flux decrease is apparent from 10--58~h.  This
  could be influenced by planetary phase variations, but the
  similarity to the purely instrumental effects seen in
  Figure~\ref{fig:young_mipscal} precludes an unambiguous distinction
  between the two effects.  }

To test this hypothesis, we look for evidence of persistence in our
data.  Using all frames taken at the second dither position we compute
the median image from each of several Astronomical Observing Requests
(AORs).  An AOR is a Spitzer logistical unit comprising some dozens of
frames; in our data set each AOR lasts approximately 3~hr.  We see
faint afterimages at the other thirteen dither positions when we
subtract the first median AOR frame from the final median AOR frame
(taken $\sim 56$~hr later;
cf. Figure~\ref{fig:crossfield_raw_timeseries}), which suggests that
the level of persistence (a byproduct of charge trapping) increases
over the course of the observations.  These afterimages are much
fainter when comparing data from the first and second AORs (separated
by 3.6~hr), consistent with the conclusion that the level of
persistence does not saturate to a constant value on these short time
scales. The afterimages are not apparent by eye when comparing the
last and penultimate AORs (again separated by 3.6~hr), which suggests
that the charge trapping persistence has saturated by this time, as
expected from the much-flattened data ramp seen in
Figure~\ref{fig:crossfield_raw_timeseries}.

The IRAC ramp is known to exhibit a behavior which depends on the
level of illumination, with more intensely illuminated pixels
exhibiting a steeper initial ramp and saturating more quickly (these
pixels' charge traps are filled more quickly because more free
photoelectrons are available).  We see a hint of this behavior in our
data.  Though pointing variations prevent us from tracking the
response of individual pixels, we extract photometry (again via PSF
fitting) using both 3- and 5-pixel-wide square apertures.  The 3-pixel
photometry -- which is weighted somewhat more heavily by the most
intensely illuminated pixels than is the 5-pixel photometry -- shows a
hint of a steeper ramp.  We take this as further tentative support for
our hypothesis that our ramp has a common origin with the IRAC ramp.
The ramp behavior remains unchanged when we use a wider aperture, but
this may not be diagnostic since the gradient in illumination level
quickly flattens out beyond a few pixels.

We would like to know why we see this ramp, especially considering
that no previous MIPS observations detected this effect. However, we
can find no consistent discriminant between the presence or absence of
a ramp in MIPS data and the state of either instrument or observatory.
The first set of AORs in \cten's observations (the first $\sim 10$~hr)
were anomalously low ($\sim0.3\%$) compared to subsequent
observations, which they attributed to a thermal anneal of the
24\,\micron\ detector conducted $<1$~h before these
observations\footnote{\protect{As recorded in the Spitzer observing
    logs, available at
    \url{http://ssc.spitzer.caltech.edu/warmmission/scheduling/observinglogs/}}}. No
ramp was observed in the continuous, long-duration MIPS observations
of either \cite{knutson:2009a} or \cten, which were taken $\gtrsim
1$~day after the last 24\,\micron\ anneal. The photometry shown in
Figure~\ref{fig:crossfield_raw_timeseries} also occurred $>1$~day after
the last 24\,\micron\ anneal, so annealing seems unlikely to explain
the presence of the ramp in our data.

We investigated whether preflashing could explain the absence of any
ramp in other MIPS phase curve observations.  To preflash is to
conduct a set of brief ($<1$~hr) observations of a bright target
before observing a fainter exoplanet system \citep{seager:2009,knutson:2011};
experience shows that this tends to reduce the amplitude of the ramp,
presumably by partially saturating the detector's charge traps.
\hdtwo\ is the faintest of the three exoplanet systems with
long-duration MIPS 24\,\micron\ observations, but the flux difference
($\sim$20~mJy for \hdtwo\ vs. $\sim60$~mJy for \hdone) does not seem
sufficiently large for only one of our five observations of \hdtwo\ to
fail to pre-flash the detector.  If the difference were due to the
increased flux from \hdone, we should still see a shorter, steeper
ramp at the start of these observations.  That no ramp has been
reported previously, and that we see a ramp in the \hdtwo\ data only
intermittently, suggests that some other phenomenon may be at work
here. 

The phase curve observations of both \hdone\ and \hdtwo\ began
immediately after a data downlink to Earth, so this factor also does
not distinguish between the cases.  Prior to the data downlinks, our
2008 observations of \hdtwo\ were preceded by 24\,\micron\
observations of the faint RXCJ0145.2-6033 ($\sim4$~mJy), but no
24\,\micron\ observations whatsoever were made in the $\sim$day
leading up to \cite{knutson:2009a}'s observations of \hdone.  While
MIPS was operational all its arrays were continuously exposed to the
sky: although the Spitzer operations staff planned observations so as
to avoid placing bright sources on the 24\,\micron\ array (using IRAS
25\,\micron\ images as a guide; A. Noriega-Crespo, private
communication) we cannot dismiss the possibility that occasionally
some bright sources may have been missed.

Thus we cannot conclusively determine why the MIPS observations of
\hdtwo\ we present here show the detector ramp while previous,
comparable observations have not shown such an effect.  Nonetheless,
the similarity between our photometry in
Figure~\ref{fig:crossfield_raw_timeseries} and raw IRAC~8\,\micron\
photometry \citep[e.g.,][]{agol:2010} strongly suggests that the most
likely explanation involves detector response variations due to
charge-trapping.

\subsection{The Fallback}
\label{sec:fallback}
After the detector ramp, the photometry in
Figure~\ref{fig:crossfield_raw_timeseries} decreases over the rest of
the observations by $\sim0.2\%$; we term this flux diminution the
``fallback.'' The amplitude of this effect is of the approximate
amplitude expected for a 24\,\micron\ planetary thermal phase curve
\citep{showman:2009,burrows:2010}, so our first inclination was to
ascribe a planetary origin to this flux decrease.  However, there is a
distinct qualitative similarity between the phase curve photometry and
pre-launch calibration data taken with the MIPS 24\,\micron\ detector
under bright (170~MJy~sr$^{-1}$) illumination, shown in
Figure~\ref{fig:young_mipscal} \citep[reproduced from][]{young:2003}.
A comparison of this figure and
Figure~\ref{fig:crossfield_raw_timeseries} reveals that both display
the same qualitative signature of an early, steep ramp followed by a
slow, gradual fallback in measured flux. The only differences are (1)
an initial steep decrease in flux in the calibration data not seen in
our stellar photometry \citep[attributed by][to the response of the
detector to a thermal anneal immediately preceding the data
shown]{young:2003}, and (2) longer ramp and fallback time constants in
our data set.

\fig{young_mipscal}{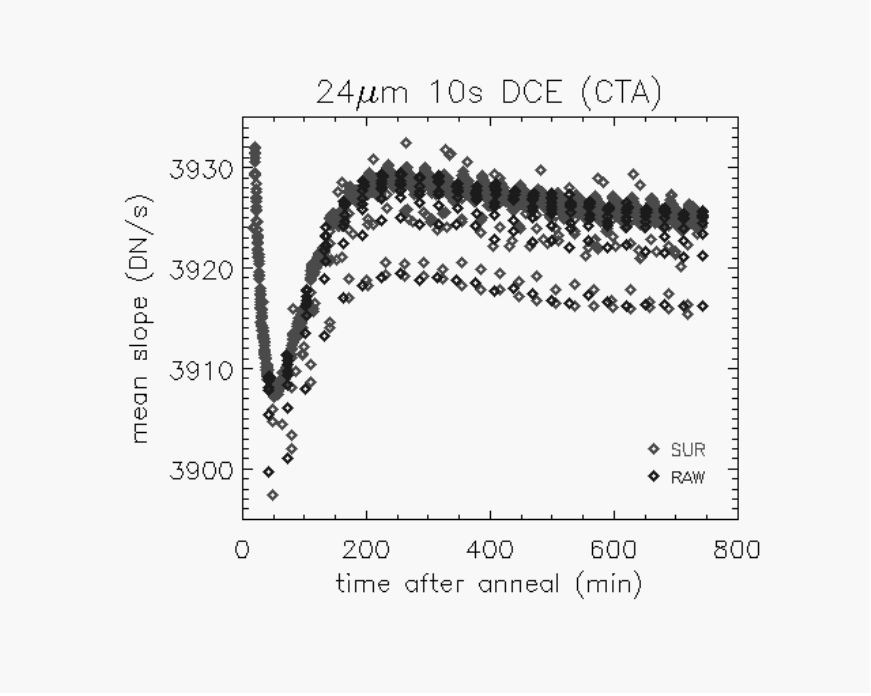}{width=3in}{}{Lab calibration
  data for the MIPS 24~\micron\ array, taken from
  \citeauthor{young:2003}~(\citeyear{young:2003}; their Figure~7). The
  relevant data for comparison with Spitzer/MIPS observations are the
  gray diamonds labeled SUR (Sample Up the Ramp, the algorithm used to
  compute MIPS data numbers from pixel slopes).  \cite{young:2003}
  suggest that the initial sensitivity decrease (0--50~minutes) is
  related to detector response variations related to a thermal anneal
  immediately preceding the data; as we describe in
  Sec.~\ref{sec:inst}, our data should not be affected by any anneal
  operations.  The rest of the observations appear strikingly similar
  to our photometry of \hdtwo, shown in
  Figure~\ref{fig:crossfield_raw_timeseries}. For comparison with
  Figure~\ref{fig:crossfield_raw_timeseries} the peak pixel fluxes in
  the \hdtwo\ data frames are roughly 1000~DN/s.}

The brightest pixels in the \hdtwo\ MIPS observations reach a flux of
45~MJy~sr$^{-1}$ (corresponding to 1000~DN~s$^{-1}$).  Perhaps, like
in some preflashed IRAC observations \citep[cf.][]{knutson:2011}, the
lower illumination level in the \hdtwo\ photometry \citep[relative to
the stimulation response curve from][]{young:2003} is responsible for
the different timescales evident in the two 24\,\micron\ time series.
However, the brightest pixels in the observations of \cten\ reached a
flux of 9000~DN~s$^{-1}$ and no fallback is apparent in the continuous
portion of those observations (though \cten's continuous photometry
did decrease monotonically by $\sim 0.1\,\%$, they demonstrated a
coherent planetary phase curve in two data sets spanning several
years: thus planetary emission, rather than an instrumental
sensitivity variation, seems a more likely interpretation of their
results).  Similarly, no fallback is seen in MIPS observations of
\hdoneb\ \citep[peak pixel flux
$\sim1200$~DN~s$^{-1}$;][]{knutson:2009a} or of the fainter eclipsing
M binary GU~Boo \citep[$\lesssim500$~DN~s$^{-1}$;][]{vonbraun:2008}.
Thus is seems possible that the fallback is linked to the presence of
the detector ramp, which also appears only in our MIPS data set.

%\subsubsection{Fitting the Fallback}
%\label{sec:fit_fallback}
We try a number of different functional forms to fit to the $\sim
0.2\%$ post-ramp fallback, which we fit simultaneously with the
ramp. These include a flat model (i.e., no decrease), sinusoidal and
Lambertian profiles with arbitrary amplitude and phase (representative
of a planetary phase curve), and a double-exponential of the form $(1
- \alpha e^{-t/\tau_1}) \times e^{-t/\tau_2}$, with $\tau_2 \gg
\tau_1$, motivated by the detector response variations seen in
Figure~\ref{fig:young_mipscal}.  We decide which of these models is
the most appropriate on the basis of the Bayesian Information
Criterion (BIC\footnote{$\textrm{BIC} = \chi^2 + k \ln N$, where $k$
  is the number of parameters to be fit and $N$ the number of data
  points.  A fit that gives a lower BIC is preferred over a fit with a
  higher BIC, and thus the BIC penalizes more complicated models. }).
The model consisting of a ramp plus a decaying exponential gives the
lowest BIC: $\sim 15$ units lower than obtained with the sinusoidal or
Lambertian models.  Thus the data prefer an instrumental explanation
for the low-level flux variations that we see.

When using a sinusoidal or Lambertian model, the best-fit phase curve
parameters describe a thermal phase variation which peaks well before
secondary eclipse, suggesting a planetary hot spot eastward of the
substellar point.  Qualitatively, such a shift is consistent with
observations of both HD~189733b \citep{knutson:2009a} and \uab\
(\cten).  However, the phase offset determined by this fitting process
is surprisingly large: $136\deg \pm 18 \deg$, a result which would seem to
imply that the planet's night side is hotter than its day side.  Such
a scenario has been predicted by some models \citep[cf.][]{cho:2003},
but such a large phase offset is bigger than observed for either \uab\
or HD~189733b, and larger still when compared to expectations for this
planet from more recent simulations
\citep[e.g.,][]{rauscher:2008,showman:2009}.  We thus deem the phase
curve fit with large offset to be an unlikely result, providing one
more reason to doubt that the flux variation we see is of planetary
origin.

We also inject into the data a sinusoidal phase curve with zero phase
offset and a peak-to-valley amplitude equal to our best-fit secondary
eclipse depth results and repeat our analysis: in this case the
best-fit sinusoidal and Lambertian models have a lower BIC value (by
12 units) than the instrumental model, though the recovered amplitude
and phase offset are still somewhat biased by the flux fallback.
Although these results suggest that we are close to achieving the
sensitivity required to constrain \hdtwob's thermal phase variations,
our ignorance of the detailed morphology of the flux fallback prevents
us from reaching a more quantitative conclusion.  Thus, we can only
conclude that the striking qualitative similarity between
Figures~\ref{fig:crossfield_raw_timeseries}
and~\ref{fig:young_mipscal} precludes us from making any definite
claims as to the detection of planetary phase curve effects in our
data.

\subsection{Instrument Stability}
As observed previously by \cten , the background flux of continuous
MIPS photometry exhibits a roughly linear trend with time, with
smaller, abrupt changes from one AOR to the next. The linear trend can
be explained by a variation in the thermal zodiacal light as Spitzer's
perspective of HD~209458 changes with respect to the solar system, and
\cten\ attribute the discontinuous, AOR-by-AOR background fluctuations
to scattered light. Whatever the cause, these discontinuities are
removed by the sky background subtraction, and do not appear to affect
the final stellar photometry.

During our 2008 observations we see a 0.5~$\mu$A increase in the
24\,\micron\ detector anneal current (MIPS data file keyword
AD24ANLI), a decrease of 6~mK in the scan mirror temperature (keyword
ACSMMTMP), and swings in the electronics box temperature (keyword
ACEBOXTM) of up to 0.3~K. During sustained observations the
electronics box appears to experience heating with some time lag, but
with a much shorter cooling lag during observational breaks to
transmit data to Earth. Upon reexamination of past observations, we
find that these three parameters exhibit similar behavior during
observations of HD~189733b \citep{knutson:2009b} and of \uandb\
\ctenp. The MIPS optical train is cryogenically cooled and separated
from the non-cryogenic instrument electronics \citep{heim:1998}, so it
does not seem likely that the observed swings in the electronics box
temperature should influence the photometry.  Similarly, the anneal
current and scan mirror temperature do not seem to correlate with
either the ramp or the post-ramp flux decrease, so we conclude that
these instrumental variations do not affect our final photometry.

\section{Transits}
\label{sec:transit}
\subsection{Fitting Approach}
We fit transits using uniform-disk and linear limb-darkened transit
models \citep{mandel:2002}, but \citep[consistent with the results
of][]{richardson:2006} we find the limb-darkened model offers no
improvement over the uniform-disk model (as determined by the BIC).
We fit the transit data for: the time of center transit $T_{c,t}$, the
impact parameter $b$, the scaled stellar radius $R_*/a$, the
planet/star radius ratio $R_p/R_*$, and the out-of-transit system flux
$F_*$.  We hold the period fixed at \fixedperiod\ \citep{torres:2008},
which is a more precise determination than our observations can
provide.  To extract useful information from our half-transit event we
always require that $b$ and $R_*/a$ have the same value, determined
jointly from all our transits.  We therefore perform one fit in which
these two parameters are jointly fit, and a second fit in which we
additionally fit jointly to $T_{c,t}$ and $R_p/R_*$ across all transit
events.

We fit to the detector ramp in the~2008 transit by including a
multiplicative factor of the form $1 - \alpha e^{-t/\tau}$, where $t$
is measured from the start of the observations.  This formulation of
the ramp model is motivated by a physical model of the charge-trapping
phenomenon thought to cause the IRAC~8\,\micron\ ramp
\citep{agol:2010}. \cite{agol:2010} find a ramp based on two
exponentials to be preferred for their high S/N observations, but we
find that our data are not precise enough to constrain this more
complicated model: when fitting a double-exponential ramp of the form
$1 - \alpha_1 e^{-t/\tau_1} - \alpha_2 e^{-t/\tau_2}$
\citep{agol:2010} the parameters for the two exponential trends become
degenerate, and the resulting fits are not preferred to the single
ramp fit on the basis of the BIC.  Finally, we include in all our fits
the fourteen sensitivity correction terms ($c_i$) corresponding to the
fourteen MIPS dither positions.

\subsection{Results}
Table~\ref{tab:joint_transit} lists the results of the fit in which we
assume a constant orbit and transit -- holding $b$, $R_*/a$,
$R_p/R_*$, and $T_{c,t}$ constant across all transits -- while
Table~\ref{tab:semijoint_transit} lists the results of the fit in
which $R_p/R_*$ and $T_{c,t}$ (but not $b$ or $R_*/a$) are allowed to
vary between events.  We plot the results of fits to each individual
transit, and to the combined data set, in
Figure~\ref{fig:jointtransit}.  We show how the residuals to the
combined fit bin down with increasing sample size in
Figure~\ref{fig:bindown}: the curve shown tracks closely with the
$N^{-1/2}$ expectation from uncorrelated noise on short time scales
($<20$~min), but on longer time scales the residuals bin down more
slowly than this.  This indicates the presence of correlated (red)
noise \citep[cf.][]{pont:2006} in these data, which is not surprising
considering the ramp residuals apparent in
Figure~\ref{fig:jointtransit}.

\figtwocol{jointtransit}{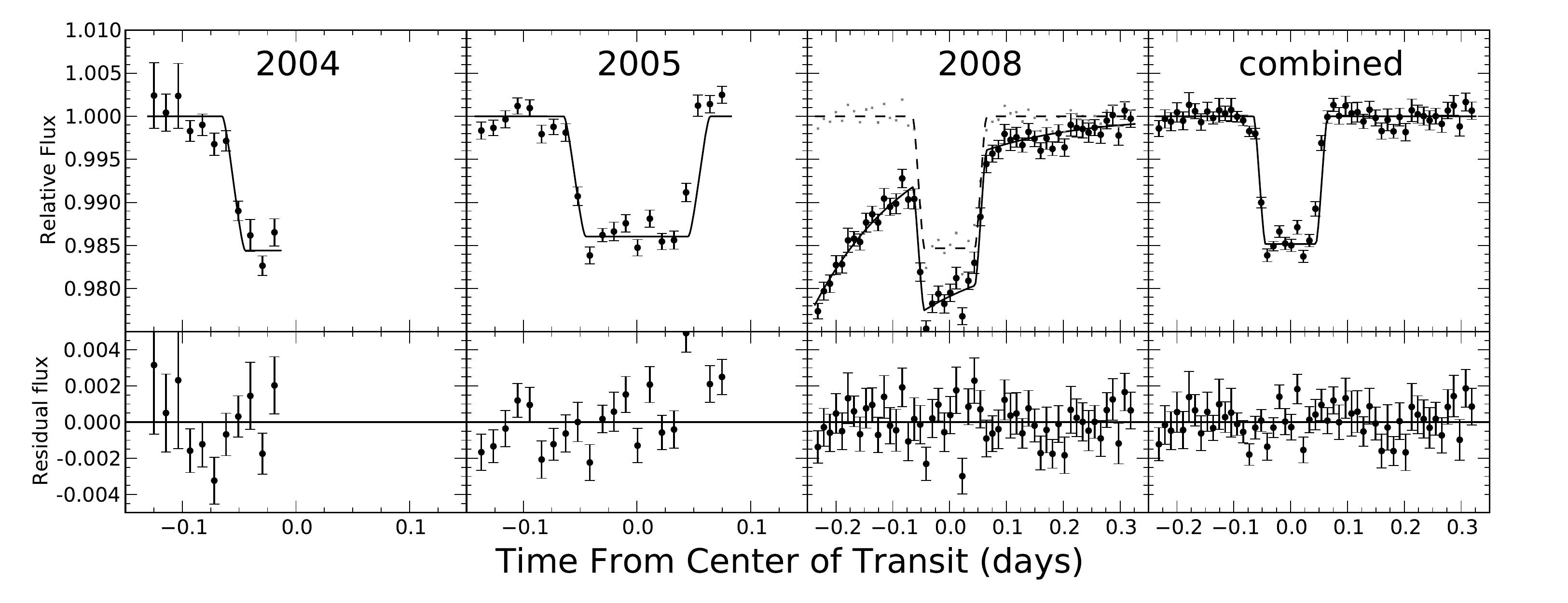}{width=6in}{}{MIPS
  24~\micron\ transits of \hdtwob.  The top panels show photometry
  and the best-fit model, and the lower panels show the residuals to
  the fits.  For plotting purposes the data have been corrected for
  the MIPS 14-position sensitivity variations, normalized by the
  stellar flux, and binned by 70 points (for the individual transits)
  and by 210 points (for the combined data set).  We also corrected
  for the ramp in the 2008 data set (corrected, binned data shown as
  small points) before combining the data to plot the data in the
  rightmost panel.  }

\fig{bindown}{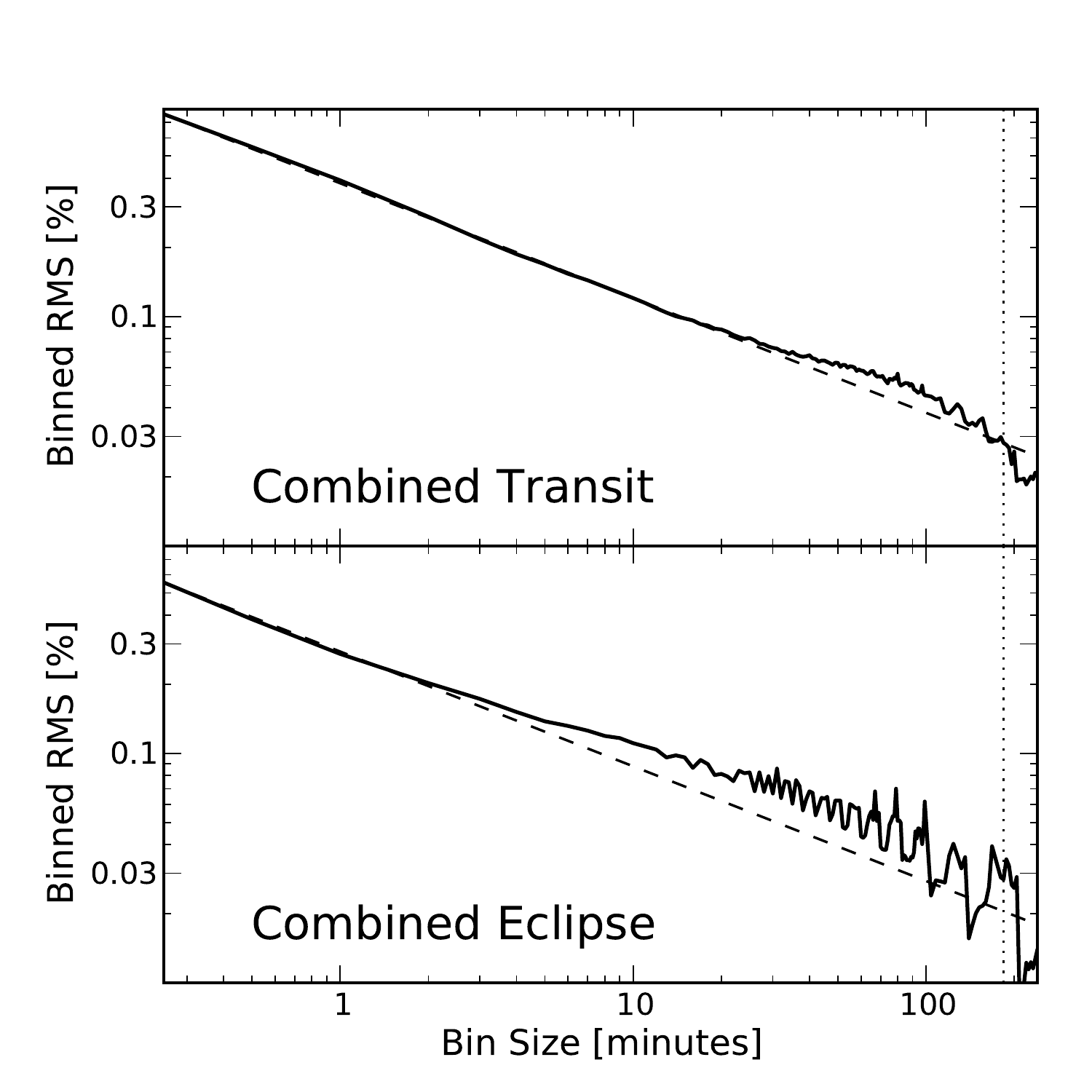}{width=3in}{}{Dispersion of the
  binned residuals (solid lines) to the combined transit and eclipse
  light curve fits shown in Figure~\ref{fig:jointtransit}
  and~\ref{fig:jointeclipse}. On longer timescales both fits exhibit a
  binned dispersion 10-30\% higher than expected from uncorrelated
  noise (dashed line).  The dashed lines show the expectation for
  uncorrelated errors, which scale as $N^{-1/2}$.  The vertical dotted
  line indicates the transit duration.}

We examine the residuals to the fourteen individual channels and see
some evidence for qualitatively different correlated noise at
different dither positions.  We do not think it likely that this
behavior is related to an intrapixel effect \citep[as observed in
IRAC; cf.][]{charbonneau:2005}, because the residual behavior we see
does not correlate with mean PSF position relative to the boundaries
of individual pixels.  Instead, it seems more likely to be a
manifestation of the known position-dependent sensitivity effect
previously attributed to residual flat-fielding errors
\citep{crossfield:2010}.

The resulting posterior distributions are all unimodal (except for the
impact parameter $b$), and the usual correlations are apparent between
$b$ and $R_*/a$ and between $F_*$ and $R_p/R_*$
\citep[cf.][]{burke:2007}.  As noted above, the 2008 transit data are
strongly affected by the detector ramp, and we see correlations
between the ramp parameters and the transit depth.  We compute the
two-dimensional posterior distributions of $R_P/R_*$, $\tau$, and
$\alpha$ (marginalized over all other parameters) from the MCMC chains
using the kernel density estimate approach described in \cten; we show
these distributions in Figure~\ref{fig:ramp_corr} and list the
elements of these parameters' covariance matrix in
Table~\ref{tab:ramp_cov}.

\fig{ramp_corr}{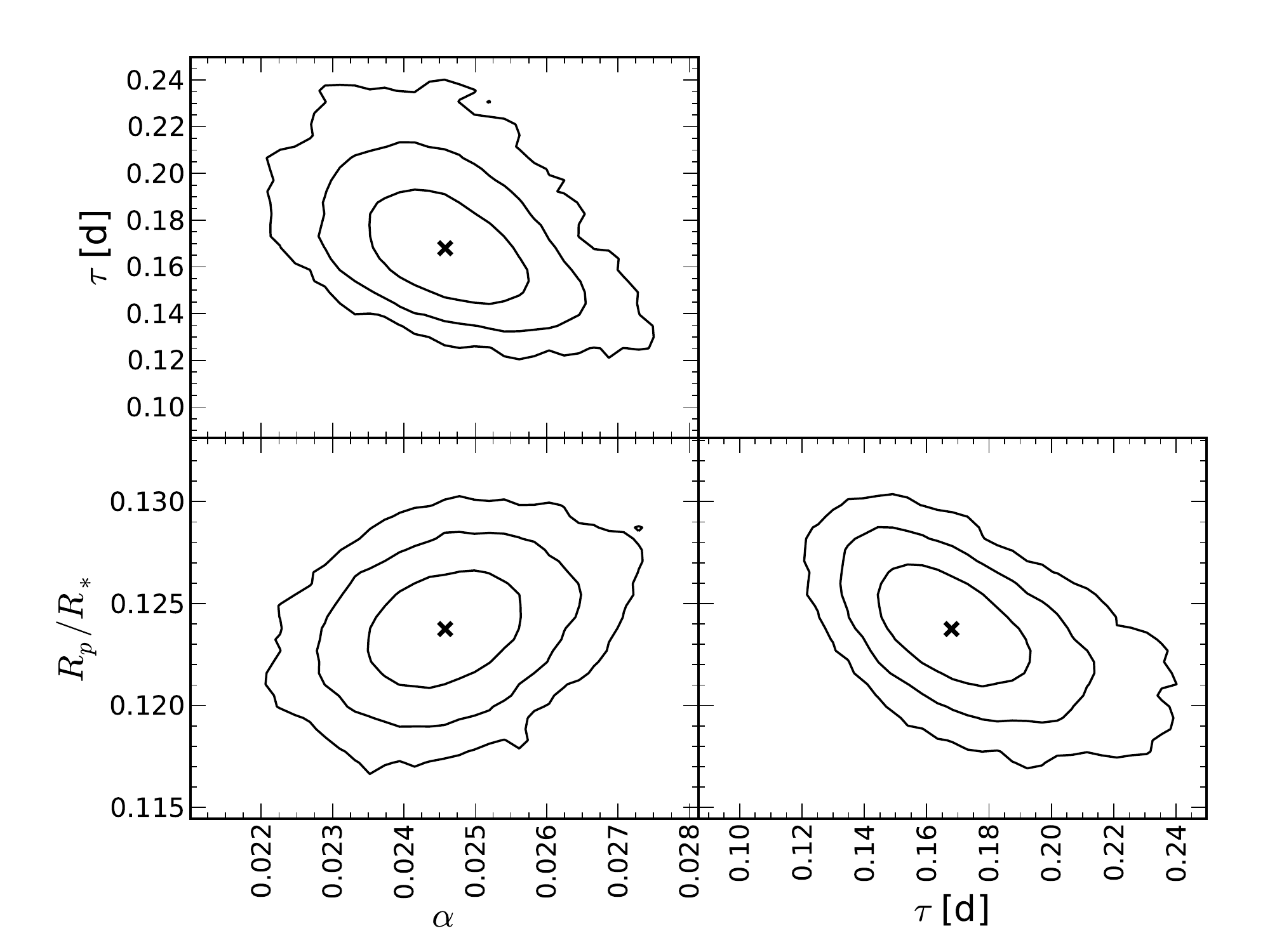}{width=3in}{}{Posterior
  distributions of the ramp parameters $(\alpha, \tau)$ and $R_p/R_*$,
  estimated from the MCMC analysis of the 2008~transit data.  The
  `$\times$' symbols indicate the best-fit parameters listed in
  Table~\ref{tab:semijoint_transit}, and the lines indicate the
  68.27\%, 95.45\%, and 99.73\% confidence intervals. The elements of
  these parameters' covariance matrix are listed in
  Table~\ref{tab:ramp_cov}.}

\subsection{Discussion}
The three independently-fit transit depths listed in
Table~\ref{tab:semijoint_transit} have a fractional dispersion of 3\%,
consistent with our individual uncertainty estimates of 3-10\%. We
thus find no evidence for variations in transit depth, and our transit
depths are consistent with the depth measured from the combination of
our first two transit data sets \citep{richardson:2006}.

We plot the ensemble of \hdtwob's transit depth measurements in
Figure~\ref{fig:transmission} along with a model of transit depth
vs. wavelength from \cite{fortney:2010}.  The model is consistent with
the 24\,\micron\ measurement we present here and agrees fairly well
with the optical measurements of \cite{sing:2008} and the IRAC~3.6
and~4.5\,\micron\ measurements of \cite{beaulieu:2010}.  However, our
model strongly disagrees with the IRAC~5.8 and~8.0\,\micron, which was also shown for the same HD 209458b model in \cite{fortney:2010}.  The large discrepancy remains unclear.  Given the known wavelength-dependent water vapor opacity, \cite{shabram:2011} showed that reaching all four 4 IRAC data points may be impossible within the framework of a simple transmission spectrum model.  Our transmission spectrum methods are described in these papers, and the atmospheric pressure-temperature profile is from a planet-wide average no-inversion model shown in Figure~\ref{fig:tpprofiles}.

\fig{transmission}{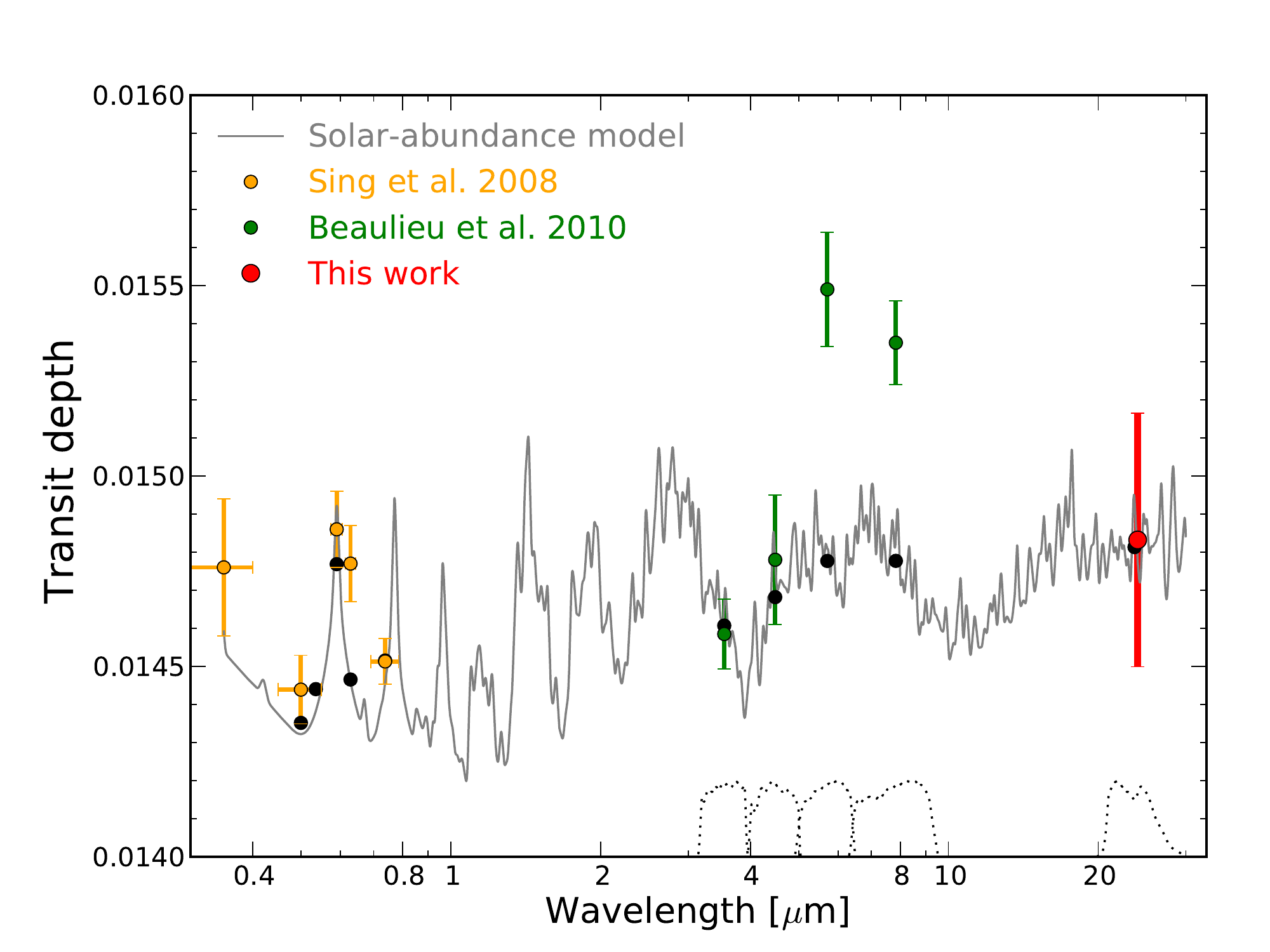}{width=3in}{}{Measurements
  of the transit depth of \hdtwob: binned optical spectroscopy
  \citep{sing:2008}, previous mid-infrared photometry
  \citep{beaulieu:2010}, and our 24\,\micron\ measurement.  The solid
  line is a model generated using the (dot-dashed)
  temperature-pressure profile shown in Figure~\ref{fig:tpprofiles}.
  The solid black points without errorbars represent the weighted
  averages of the model over the corresponding bandpasses (indicated
  at bottom).}

We resample the posterior distributions of the independent transit
ephemerides shown in Table~\ref{tab:semijoint_transit} to determine
our own, independent constraint on the planet's orbital period
(assuming it is constant) using a linear relation. We compute the
center-of-transit time and period to be \fittransitephemeris\ and
\fittransitperiod, respectively; the covariance between these two
parameters is \fittransitcov. The period we obtain differs from the
established period \citep{torres:2008} by only \fittransitdifference\
(\fittransitdifferences), well within the uncertainties.

\fig{tpprofiles}{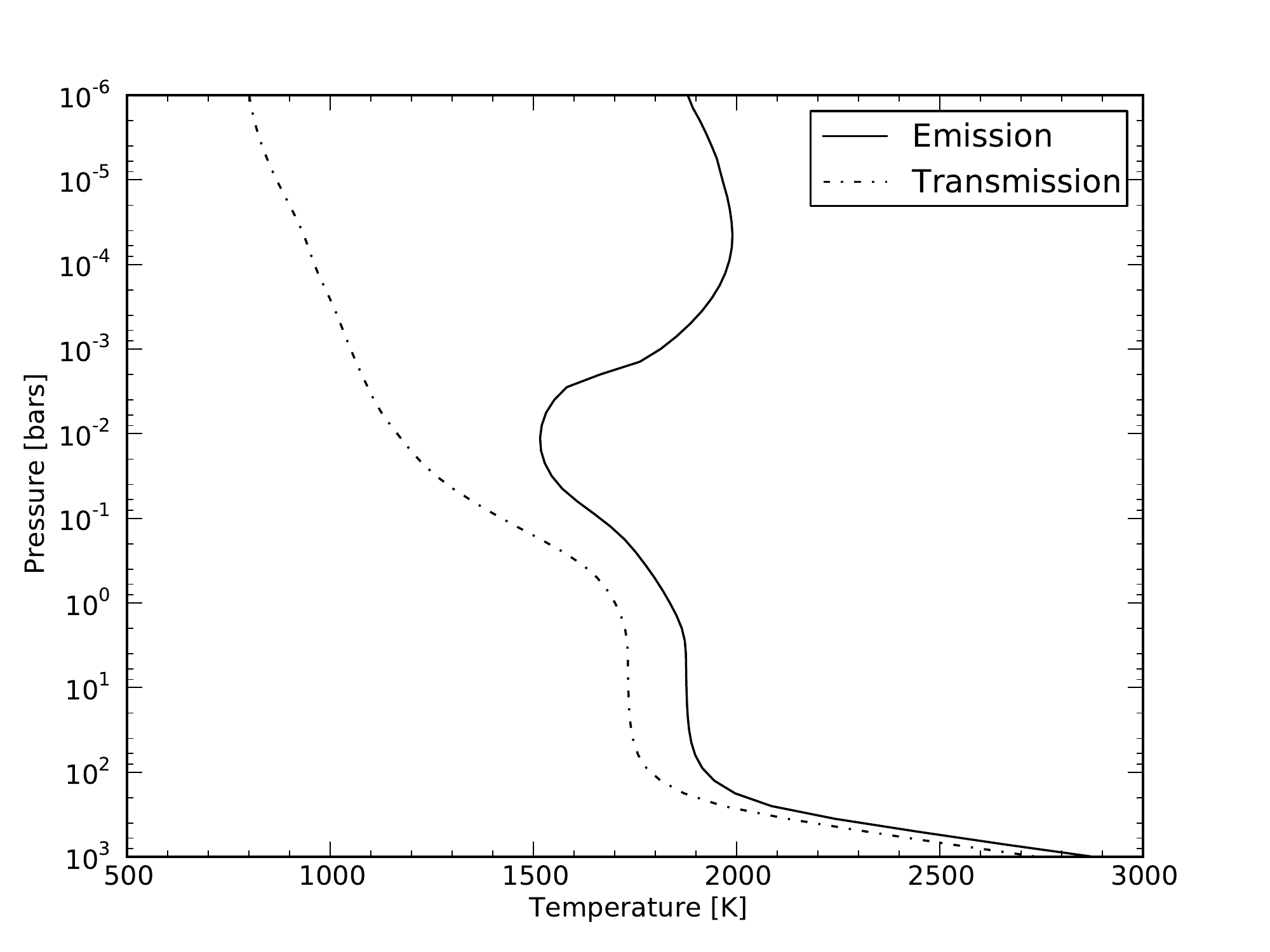}{width=3in}{}{Temperature-pressure
  (T-P) profiles used to generate our model spectra.  The dot-dashed
  curve is a planet-wide average T-P profile taken from a full (4$\pi$)
  redistribution model, and is used to model the tranmission spectrum
  shown in Figure~\ref{fig:transmission}. It includes TiO/VO opacity,
  but these species have only a minor effect since nearly all of the Ti/V has
  condensed out of the gas phase at these cooler temperatures. The
  solid curve is from a model assuming no redistribution of absorbed
  energy (making it hotter), and includes TiO/VO to drive the
  temperature inversion seen in Figure~\ref{fig:emission}.}

\section{Secondary Eclipses}
\label{sec:eclipse}
\subsection{Fitting Approach}
We fit secondary eclipses using the uniform-disk occultation formulae
of \cite{mandel:2002}, fitting each event for three astrophysical
parameters: time of center of eclipse $T_{c,e}$, stellar flux $F_*$,
and eclipse depth $F_p/F_*$ -- as well as the fourteen sensitivity
correction terms ($c_i$) discussed previously. We hold all other other
orbital parameters fixed at the values listed in \cite{torres:2008},
which are more precise than our constraints based on the 24\,\micron\
transit photometry.  We perform four different fits: an independent
fit of each eclipse taken in isolation, and a fit to the combined data
set in which we fit for a single eclipse depth, but still allow
$T_{c,e}$ and $F_*$ to vary for each event.  We use only a subset of
the long-duration phase curve observations to fit the 2008 eclipse, as
indicated in Table~\ref{tab:obs}. We tried including a linear slope in
the combined eclipse fit, but this extra parameter is not justified
because it gives a higher BIC than fits without such a slope.

\subsection{Results}
The parameters for the fit in which $T_{c,e}$ and $F_p/F_*$ are fit
jointly across all eclipses (but $F_*$ remains independent) are shown
in Table~\ref{tab:jointeclipse}, and parameters for the three wholly
independent eclipse fits are shown in Table~\ref{tab:indepeclipse}.
The data, best fit models, and residuals for all three eclipses and
the combined data set are plotted in Figure~\ref{fig:jointeclipse}.
The only strong correlations apparent in the resulting posterior
distributions are between $F_*$ and $F_p/F_*$ -- expected since we are
making a relative measurement.  We show how the residuals to the
combined fit bin down with increasing sample size in
Figure~\ref{fig:bindown}: the residuals average down more slowly than
the $N^{-1/2}$ expectation from uncorrelated errors.  This indicates
the presence of correlated (red) noise \citep[cf.][]{pont:2006} in
these data, which is expected given the behavior of the eclipse
residuals shown in Figure~\ref{fig:jointeclipse}.

\figtwocol{jointeclipse}{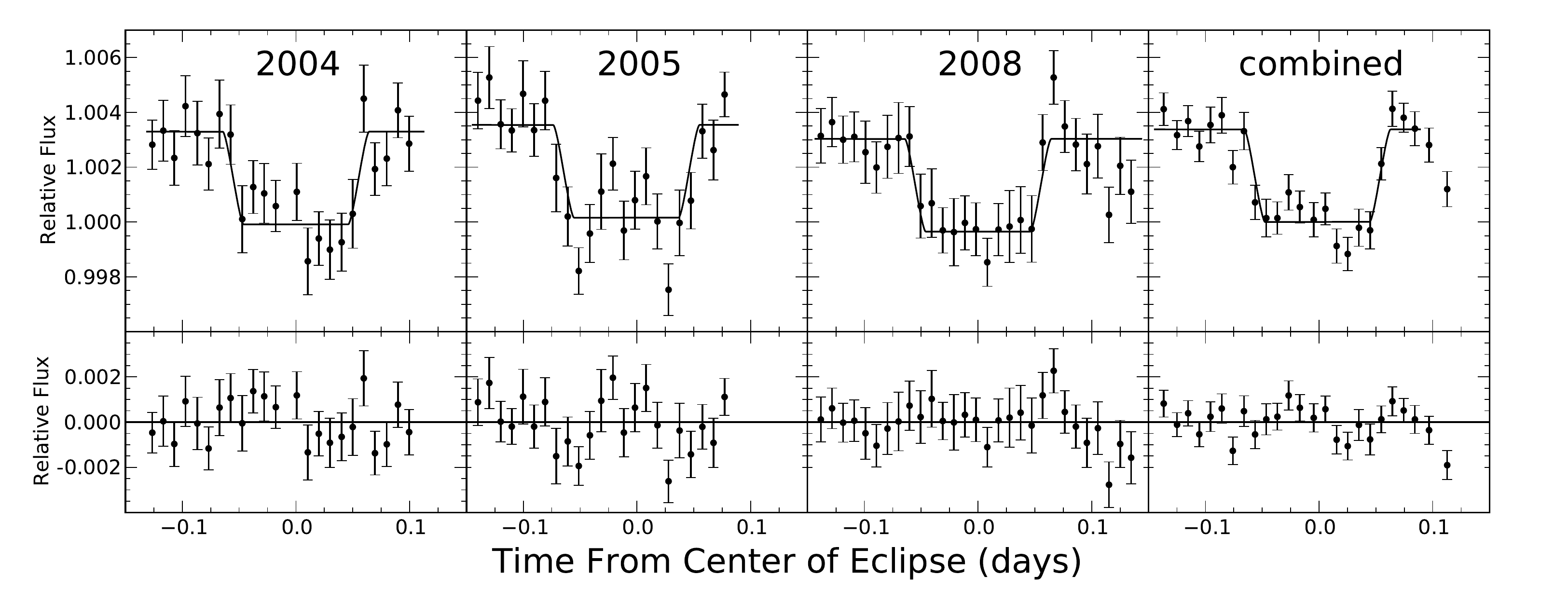}{width=6in}{}{MIPS
  24~\micron\ eclipses of \hdtwob.  The top panels show photometry and
  the best-fit models, and the lower panels show the residuals to the
  fits. For plotting purposes the data have been corrected for the
  MIPS 14-position sensitivity variations, normalized by the stellar
  flux, and binned by 70 points (for the individual eclipses) and by
  210 points (for the combined data set).  }

\subsection{Discussion}
The three eclipse depths have a dispersion of 13\%, consistent with
our estimated measurement errors (12-18\%).  We thus find no evidence
for variability of planetary emission, in good agreement with general
circulation models which predict \hdtwob's MIR dayside emission will
vary by $<5\%$
\citep[e.g.,][]{rauscher:2008,showman:2009,dobbs-dixon:2010} and
consistent with the measurement that \hdoneb's 8\,\micron\ dayside
emission varies by $<2.7\,\%$\citep{agol:2010}.  Our mean eclipse
depth over all three epochs -- \jeclipses\ -- is $\sim 1.3\sigma$
deeper than the initial measurement by \cite{deming:2005a} of $0.26\%
\pm 0.046\%$.  We convert this eclipse depth to a brightness
temperature of \daysidetemps\ using the method outlined by \cten.

We plot the ensemble of \hdtwob's secondary eclipse measurements in
Figure~\ref{fig:emission} along with a model of planet/star
contrast ratio vs. wavelength.  The modeling procedure is described in detail in \cite{fortney:2006} and \cite{fortney:2008}.  Using a stellar model for the incident flux and a solar metallicity atmosphere, we derive a radiative-convective pressure-temperature profile assuming chemical equilibrium mixing ratios.  The model assumes no loss of absorbed energy to the night side, and redistribution of energy over the day side only \citep[see ][]{fortney:2008}.  We show the pressure-temperature profile, which feature a temperature inversion due to the absorption of stellar flux by TiO and VO gasses, in Figure~\ref{fig:tpprofiles}.  Clearly a stronger temperature inversion is needed, as the contrast between the IRAC 3.6 and 4.5 \,\micron\ bands is not large enough.  Since the 24\,\micron\
photosphere is predicted to lie at $1-10$~mbar on \hdtwob\
\citep{showman:2009} our measurement indicates a somewhat cooler
temperature than is expected for this planet given its atmospheric
temperature inversion.  The anomalously low 24\,\micron\ flux has been
noted previously \citep[e.g.,][]{madhusudhan:2010}; taken in concert
with \uab 's large and still-unexplained 24\,\micron\ phase offset
(\cten) these results suggest that our current understanding of
atmospheric opacity sources in this wavelength range may be
incomplete.  Alternatively, we can fit reasonably fit the 3.6, 8.0, and 24 \,\micron\ points with the dayside emission of the 3D general circulation model of \cite{showman:2009}, which is cooler than the corresponding 1D model from \cite{fortney:2008}.  Clearly more work is needed to robustly fit the dayside photometry of the planet within the framework of a 1D or 3D self-consistent model.

\fig{emission}{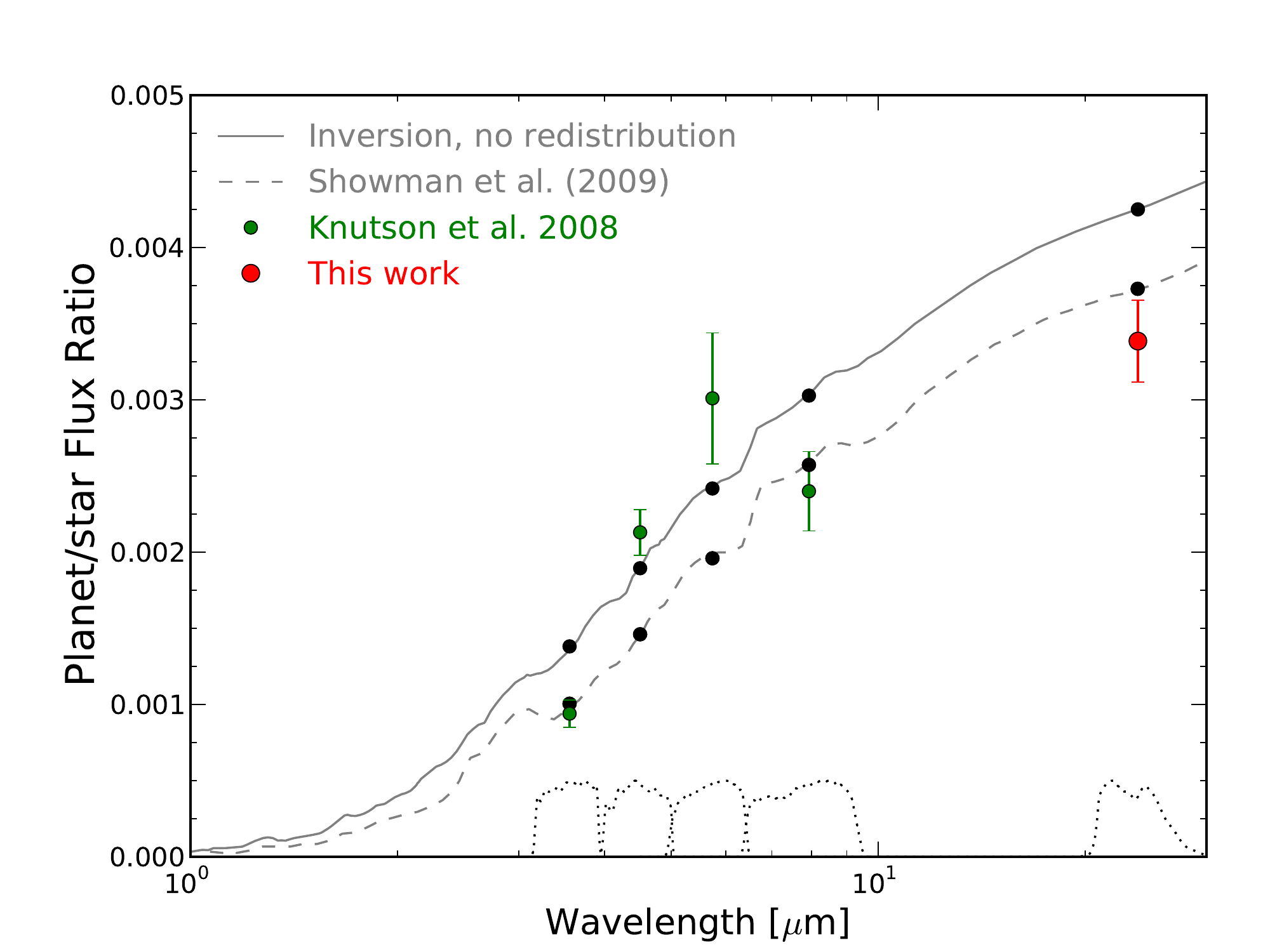}{width=3in}{}{Measurements
  of the secondary eclipse depth of \hdtwob: previous Spitzer/IRAC
  photometry \citep{knutson:2008} and our 24\,\micron\ measurement.
  The solid line is from a model assuming zero redistribution of
  incident flux and including gaseous TiO and VO to drive a
  temperature inversion; we show this model's temperature-pressure
  profile in Figure~\ref{fig:tpprofiles}.  The dashed line is the
  emission spectrum from \cite{showman:2009}. The solid black points
  without errorbars represent the weighted averages of the models over
  the corresponding bandpasses (indicated at bottom).}

We also fit a linear relation to the three eclipse times in the same
manner as in Sec.~\ref{sec:transit}.  We compute a period of
\fiteclipseperiod, which differs from the established period
\citep{torres:2008} by \fiteclipsedifference\
(\fiteclipsedifferences), well within the uncertainties.  This value
also agrees with our measurement of the period from the transit fits;
the two periods differ by only \timediff, which is (as expected)
consistent with zero.

\section{Joint Orbital Constraints and System Flux}
\label{sec:misc}
\subsection{Timing and Eccentricity: Still a Chance for Winds}
Measuring the times of transit and secondary eclipse constrains the
quantity $e \cos \omega$, where $e$ is the planet's orbital
eccentricity and $\omega$ its longitude of periastron \citep[][chapter
by J.~Winn]{seager:2011}. We resample the posterior distributions of
$T_{c,t}$ and $T_{c,e}$ from the fits shown in
Tables~\ref{tab:joint_transit} and~\ref{tab:jointeclipse} and compute
the difference between our transit and eclipse ephemerides (i.e.,
$\left[ T_{c,e} - T_{c,t} \right] \textrm{~mod~} P - \frac{P}{2}$) to
be \deltatc\ after also accounting for the 47~s light travel time from
the planet's location during eclipse to its location during transit
\citep{torres:2008}. This results constrains $e \cos \omega$ to be
\ecosomega, consistent with zero and with previous constraints from
radial velocity \citep{torres:2008}.  We do not see the marginal
timing offset previously reported \citep{knutson:2008}, which may have
been biased by the higher level of correlated noise (due to the IRAC
intrapixel effect) in the~3.6 and~4.5\,\micron\ IRAC data.

A measurement of $e \cos \omega$ directly constrains the apparent velocity
offset that can be induced in planetary absorption lines during
transit \cite[cf.][]{montalto:2011}; this provides an independent
check as to whether the recent measurement of a velocity offset of $2
\pm 1 \textrm{~km~s}^{-1}$ in \hdtwob\ \citep{snellen:2010} can be
attributed to a low, but nonzero, orbital eccentricity.  Our timing
measurements of \hdtwob\ set a $3\sigma$ upper limit on any velocity
offset due to the planet's orbital eccentricity of only $140
\textrm{~m~s}^{-1} / \sqrt{1 - e^2}$.  Thus the claimed velocity
offset, though still of low significance, cannot be dismissed as
resulting from the \hdtwob's orbital eccentricity.

\subsection{System Flux: No Excess Detected}
Although our primary science results -- the transit and eclipse depths
-- rely on relative flux measurements, our observations also allow us
to measure absolute 24\,\micron\ photometry for the \hdtwo\ system.
Our flux measurements for this system vary from epoch to epoch by much
more than our quoted statistical uncertainties, but the variations are
not large compared to the $\lesssim 1\%$ repeatability and 2\%
absolute calibration accuracy of the MIPS 24\,\micron\ array
\citep{engelbracht:2007}. Our 21~pixel aperture encloses 99.2\%\ of
the starlight (as determined from our synthetic PSF), and we account
for this small effect in the value quoted below.

We therefore report the 24\,\micron\ system flux as \systemfluxes,
consistent with the flux expected from the \hdtwo\ stellar photosphere
\citep[as reported by ][]{deming:2005a}.  \hdtwo\ was not detected by
IRAS \citep{beichman:1988}, but is present in the Widefield Infrared
Survey Explorer's all-sky point source catalogue \citep{wright:2010}.
The WISE photometry gives a W4 system flux of $25.74 \pm 0.12$~mJy,
which is higher than but marginally ($\sim 3 \sigma$) consistent with
the Spitzer-derived value after accounting for the different
wavelengths of the two instruments.  We therefore conclude that
\hdtwo\ does not have a strong 24\,\micron\ infrared excess, as is
typical of middle-aged F dwarfs \citep{moor:2011}.

\section{Conclusions and Future Work}
\label{sec:conclusion} We have described a homogeneous analysis of all
Spitzer MIPS observations of the hot Jupiter \hdtwob.  The data
comprise three eclipses, two and a half transits, and a long,
continuous observation designed to observe the planet's thermal phase
curve; of these, analysis of two of the eclipses, one transit, and the
phase curve observations have remained unpublished until now.  The
long-duration phase curve observation exhibits a detector ramp that
appears similar to the ramp seen in Spitzer/IRAC~8\,\micron\
photometry, and we model this effect using the exponential function
proposed by \cite{agol:2010}.  We also see a $\sim-0.2\%$ flux
decrease in the latter portion of the phase curve observations.  This
fallback is similar to a known (but poorly characterized) variation in
the response of the MIPS detector when subjected to bright
illumination \citep[cf. Figure~\ref{fig:young_mipscal}
and][]{young:2003}.

We are unable to determine why either the fallback or the ramp have
not been seen in any prior MIPS observations.  Despite this failure
the correspondence between our photometry and the pre-launch array
calibration data leads us to conclude that ramp and fallback are
correlated and both are most likely of instrumental, rather than
astrophysical, origin.  This conclusion is strengthened by the result
that fitting periodic phase functions to the data yields a planetary
model hotter on its night side than its day side, strikingly at odds
with theory
\citep{rauscher:2008,showman:2009,dobbs-dixon:2010,cowan:2011circ} and
inconsistent with other published observations of hot Jupiters
\citep{harrington:2006,knutson:2007b,knutson:2009a,crossfield:2010}.

We see no evidence for variation in the three eclipse depths, and a
joint fit of all three eclipses gives our best estimate of the
24\,\micron\ planet/star contrast: \jeclipses.  This value is more
precise and higher than the previously published measurement
\citep{deming:2005a}, and corresponds to an average dayside brightness
temperature of \daysidetemps, consistent with models of this planet's
thermal atmospheric structure
\citep{showman:2009,madhusudhan:2010,moses:2011}.  We note
parenthetically that this new eclipse depth has already diffused into
several papers
\citep[cf.][]{showman:2009,burrows:2010,madhusudhan:2009,madhusudhan:2010,fortney:2010,moses:2011};
the value and uncertainty quoted in those works are close to those we
report here, so their conclusions should be relatively unaffected.

We see no evidence for variations in our transit measurements, and a
joint fit of our two and a half transits yields a 24\,\micron\ transit
depth of \jtransits.  The transit depth is less well-constrained than
the eclipse depth because only half of the first transit was observed,
and the last transit occurred during the detector ramp.

The ephemerides calculated from our analyses of the transits and
eclipses allow us to compute orbital periods of \fittransitperiod\ and
\fiteclipseperiod, respectively, which are consistent with but less
precise than the orbital period of \cite{torres:2008}.  Eclipses occur
\deltatc\ earlier than would be expected from a circular orbit, which
constrains the orbital quantity $e \cos \omega$ to be \ecosomega.
This suggests that \hdtwob's inflated radius \citep[larger than
predicted by models of planetary interiors;][]{fortney:2007} cannot be
explained by interior heating from ongoing tidal circularization, and
that the possible velocity offset reported by \cite{snellen:2010}
cannot be explained by a nonzero orbital eccentricity.

Although we obtain improved estimates of the 24\,\micron\ transit and
secondary eclipse parameters, instrumental effects prevent a
conclusive detection of the planet's thermal phase curve. The phase
curve signal is inextricably combined with the systematic fallback
effect, despite estimates that the planet's day/night contrast should
be as large as a few parts per thousand \citep{showman:2009}.  Such a
large and intermittent systematic effect has profound implications for
future mid-infrared exoplanet observations with EChO, SPICA, and JWST.
Models of terrestrial planet phase curves predict phase amplitudes of
$\lesssim 10^{-4}$ \citep{selsis:2011,maurin:2011}; such observations
could be utterly confounded by the instrumental systematics seen in
our observations, and so may be much more challenging that has been
heretofore assumed \citep{kaltenegger:2009,seager:2009}.  Although it
may be possible to reduce the effect of the ramp with a pre-flash
strategy similar to that adopted for the 8\,\micron\ IRAC array, a
further defense against these challenges would seem to be a more
comprehensive campaign of array characterization. Specifically, a
detailed characterization of the detector response to sustained levels
of the high illumination expected from observations of terrestrial
planets around the brightest nearby stars is highly desirable, and
should be considered an essential requirement for all future infrared
space telescopes.

\acknowledgments

We thank Brad Hansen for many informative discussions, and Alberto
Noriega-Crespo and James Colbert of the Spitzer Science Center for
discussions about calibration of, and systematics in, MIPS
24\,\micron\ photometry. We thank the referee for useful comments and
the suggestion to expand our discussion of the residual noise properties.

This work is based on observations made with the Spitzer Space
Telescope, which is operated by the Jet Propulsion Laboratory,
California Institute of Technology under a contract with NASA.
Support for this work was provided by NASA through an award issued by
JPL/Caltech. We received free software and services from SciPy,
Matplotlib, and the Python Programming Language.  This research made
use of Tiny Tim/Spitzer, developed by John Krist for the Spitzer
Science Center; the Center is managed by the California Institute of
Technology under a contract with NASA.

\footnotesize

\bibliographystyle{apj_hyperref}

%\bibliography{ms}

\clearpage

\begin{deluxetable}{lrrcrrr}
\tabletypesize{\scriptsize}
\tablecaption{\emph{Spitzer}/MIPS 24\,\micron\ Observations of HD 209458b \label{tab:obs}}
\tablewidth{0pt}
\tablehead{
\colhead{UT Date} & \colhead{Event} & \colhead{Duration (hr)} & \colhead{t$_{int}$ (s)} & \colhead{N$_{exposures}$} & \colhead{Bkd (MJy Sr$^{-1}$)\tablenotemark{a}} & \colhead{$\Delta t$ (s)\tablenotemark{b}}}
\startdata
UT 2004 Dec 5  & Half transit & 2.8 & 8.91 & 840   & 28.8 & -544\\
UT 2004 Dec 6  & Eclipse      & 5.8 & 9.96 & 1680  & 29.2 & -531\\
UT 2005 Jun 27 & Transit      & 5.6 & 9.96 & 1680  & 28.9 & -612\\
UT 2005 Dec 1  & Eclipse      & 5.6 & 9.96 & 1680  & 26.7 & +183\\
UT 2008 Jul 25\tablenotemark{c} & Transit\tablenotemark{d} & 14.2 & 9.96 & 4060 & 28.3 & -649\\
UT 2008 Jul 27\tablenotemark{c} & Eclipse & 6.9  & 9.96 & 2072 & 27.9 & -663\\

\enddata
\tablenotetext{a}{Average sky backgrounds as reported by DRIBKGND
  keyword.}  
\tablenotetext{b}{For each event, $\Delta t \equiv \langle
  \textrm{HJD} \rangle - \langle BJD_{TDB} \rangle$}
\tablenotetext{c}{These events were observed as part of a
  single, continuous phase curve observation with a duration of 58
  hours spanning one transit and one secondary eclipses.}
\tablenotetext{d}{This transit was corrupted by an apparent ramp in
  detector sensitivity, so we used a longer section of data to better
  constrain the ramp parameters in the joint fit.}
\end{deluxetable}

\begin{deluxetable}{l c c c}
\tabletypesize{\scriptsize}
\tablewidth{0pt}
\tablecaption{\label{tab:joint_transit} Joint Transit Fits}
\tablehead{
\colhead{Parameter} & \colhead{2004} & \colhead{2005} & \colhead{2008} \\
} 
\startdata
% Updated 2011-12-19
      $c_{0}$  &       +0.0009  $\pm$ 0.0022  &       +0.00036  $\pm$ 0.00086  &       +0.00274  $\pm$ 0.00053  \\
      $c_{1}$  &       +0.0097  $\pm$ 0.0027  &       +0.00881  $\pm$ 0.00088  &       +0.01387  $\pm$ 0.00055  \\
      $c_{2}$  &       -0.0030  $\pm$ 0.0022  &       +0.00101  $\pm$ 0.00136  &       +0.00487  $\pm$ 0.00053  \\
      $c_{3}$  &       +0.0106  $\pm$ 0.0024  &       +0.00866  $\pm$ 0.00079  &       +0.00858  $\pm$ 0.00056  \\
      $c_{4}$  &       -0.0034  $\pm$ 0.0024  &       +0.00148  $\pm$ 0.00115  &       -0.00020  $\pm$ 0.00053  \\
      $c_{5}$  &       -0.0021  $\pm$ 0.0032  &       +0.01006  $\pm$ 0.00080  &       +0.01310  $\pm$ 0.00052  \\
      $c_{6}$  &       +0.0017  $\pm$ 0.0022  &       -0.00574  $\pm$ 0.00078  &       -0.00502  $\pm$ 0.00057  \\
      $c_{7}$  &       -0.0023  $\pm$ 0.0033  &       +0.00167  $\pm$ 0.00081  &       -0.00720  $\pm$ 0.00054  \\
      $c_{8}$  &       -0.0036  $\pm$ 0.0023  &       -0.00298  $\pm$ 0.00078  &       -0.00121  $\pm$ 0.00058  \\
      $c_{9}$  &       +0.0098  $\pm$ 0.0023  &       -0.00324  $\pm$ 0.00135  &       -0.00919  $\pm$ 0.00060  \\
     $c_{10}$  &       -0.0032  $\pm$ 0.0028  &       +0.00394  $\pm$ 0.00111  &       +0.00210  $\pm$ 0.00057  \\
     $c_{11}$  &       -0.0126  $\pm$ 0.0032  &       -0.00974  $\pm$ 0.00103  &       -0.01016  $\pm$ 0.00057  \\
     $c_{12}$  &       +0.0008  $\pm$ 0.0028  &       -0.00039  $\pm$ 0.00085  &       -0.00122  $\pm$ 0.00065  \\
     $c_{13}$  &       -0.0028  $\pm$ 0.0024  &       -0.01360  $\pm$ 0.00091  &       -0.01062  $\pm$ 0.00052  \\
  $F_*$ [mJy]  &      +18.845  $\pm$ 0.012  &      +18.7784  $\pm$ 0.0049  &      +18.696  $\pm$ 0.010  \\
     $\alpha$  & --  & --   &       +0.02437  $\pm$ 0.00068  \\
   $\tau$ [d]  & --  & --   &       +0.174  $\pm$ 0.016  \\
\hline
\multicolumn{2}{r}{$T_{c,t}$ [BJD$_{TDB}$]} & \multicolumn{2}{l}{ 2453549.20852  $\pm$ 0.00049}  \\
\multicolumn{2}{r}{          $b$} & \multicolumn{2}{l}{       +0.590  $\pm$ 0.062}  \\
\multicolumn{2}{r}{      $R_*/a$} & \multicolumn{2}{l}{       +0.1205  $\pm$ 0.0066}  \\
\multicolumn{2}{r}{    $R_p/R_*$} & \multicolumn{2}{l}{       +0.1218  $\pm$ 0.0014}  \\
\multicolumn{2}{r}{$(R_p/R_*)^2$\tablenotemark{a}} & \multicolumn{2}{l}{       +0.01483  $\pm$ 0.00033}  \\

\enddata
\tablenotetext{a}{Computed from the posterior MCMC distributions of $R_P/R_*$.}
\end{deluxetable}

\begin{deluxetable}{l c c c}
\tabletypesize{\scriptsize}
\tablewidth{0pt}
\tablecaption{\label{tab:semijoint_transit} Semi-Joint Transit Fits}
\tablehead{
\colhead{Parameter} & \colhead{2004} & \colhead{2005} & \colhead{2008} \\
} 
\startdata
% Updated 2011-11-05

      $c_{0}$  &       -0.0001  $\pm$ 0.0024  &       +0.00043  $\pm$ 0.00084  &       +0.00268  $\pm$ 0.00054  \\
      $c_{1}$  &       +0.0091  $\pm$ 0.0030  &       +0.00866  $\pm$ 0.00095  &       +0.01368  $\pm$ 0.00053  \\
      $c_{2}$  &       -0.0032  $\pm$ 0.0022  &       +0.00087  $\pm$ 0.00149  &       +0.00524  $\pm$ 0.00076  \\
      $c_{3}$  &       +0.0117  $\pm$ 0.0023  &       +0.00872  $\pm$ 0.00082  &       +0.00878  $\pm$ 0.00066  \\
      $c_{4}$  &       -0.0032  $\pm$ 0.0023  &       +0.00146  $\pm$ 0.00110  &       -0.00016  $\pm$ 0.00053  \\
      $c_{5}$  &       -0.0021  $\pm$ 0.0029  &       +0.01009  $\pm$ 0.00078  &       +0.01316  $\pm$ 0.00053  \\
      $c_{6}$  &       +0.0019  $\pm$ 0.0022  &       -0.00603  $\pm$ 0.00084  &       -0.00516  $\pm$ 0.00053  \\
      $c_{7}$  &       -0.0019  $\pm$ 0.0039  &       +0.00157  $\pm$ 0.00083  &       -0.00686  $\pm$ 0.00058  \\
      $c_{8}$  &       -0.0039  $\pm$ 0.0023  &       -0.00301  $\pm$ 0.00078  &       -0.00110  $\pm$ 0.00063  \\
      $c_{9}$  &       +0.0095  $\pm$ 0.0022  &       -0.00317  $\pm$ 0.00144  &       -0.00910  $\pm$ 0.00057  \\
     $c_{10}$  &       -0.0032  $\pm$ 0.0029  &       +0.00378  $\pm$ 0.00093  &       +0.00189  $\pm$ 0.00052  \\
     $c_{11}$  &       -0.0124  $\pm$ 0.0028  &       -0.00942  $\pm$ 0.00082  &       -0.00977  $\pm$ 0.00056  \\
     $c_{12}$  &       +0.0007  $\pm$ 0.0028  &       -0.00035  $\pm$ 0.00086  &       -0.00189  $\pm$ 0.00129  \\
     $c_{13}$  &       -0.0026  $\pm$ 0.0022  &       -0.01330  $\pm$ 0.00078  &       -0.01096  $\pm$ 0.00061  \\
$T_{c,t}$ [BJD$_{TDB}$]  &  2453344.7718  $\pm$ 0.0025  &  2453549.20746  $\pm$ 0.00065  &  2454673.60391  $\pm$ 0.00074  \\
    $R_p/R_*$  &       +0.1227  $\pm$ 0.0060  &       +0.1189  $\pm$ 0.0020  &       +0.1238  $\pm$ 0.0019  \\
  $F_*$ [mJy]  &      +18.850  $\pm$ 0.016  &      +18.7735  $\pm$ 0.0056  &      +18.6947  $\pm$ 0.0095  \\
$(R_p/R_*)^2$\tablenotemark{a}  &       +0.0151  $\pm$ 0.0015  &       +0.01413  $\pm$ 0.00046  &       +0.01531  $\pm$ 0.00046  \\
     $\alpha$  & --  &  --  &       +0.02458  $\pm$ 0.00071  \\
   $\tau$ [d]  & --  &  --  &       +0.168  $\pm$ 0.016  \\
\hline
\multicolumn{2}{r}{          $b$} & \multicolumn{2}{l}{      +0.581  $\pm$ 0.070}  \\
\multicolumn{2}{r}{      $R_*/a$} & \multicolumn{2}{l}{      +0.1197 $\pm$ 0.0069}  
\enddata
\tablenotetext{a}{Computed from the posterior MCMC distributions of $R_P/R_*$.}
\end{deluxetable}

\begin{deluxetable}{l c}
\tablewidth{0pt}
\tabletypesize{\scriptsize}
\tablecaption{Ramp and transit depth covariance
  matrix \label{tab:ramp_cov} (cf. Figure~\ref{fig:ramp_corr}).}
\tablehead{
  \colhead{Element}  & \colhead{$\textrm{Value} / 10^{6}$}
}
\startdata
 % updated 2011-11-05
     $\sigma^2_{\alpha}$ &     0.507 \\ 
   $\sigma^2_{\tau [d]}$ &   266~d$^2$ \\ 
    $\sigma^2_{R_p/R_*}$ &     3.49 \\ 
$\sigma_{\alpha,\tau [d]}$ &    -5.03~d \\ 
$\sigma_{\alpha,R_p/R_*}$ &     0.379 \\ 
$\sigma_{\tau [d],R_p/R_*}$ &   -16.3~d \\ 

\enddata

\end{deluxetable}

\begin{deluxetable}{l c c c}
\tablewidth{0pt}
\tabletypesize{\scriptsize}
\tablecaption{Joint Eclipse Fits \label{tab:jointeclipse}}
\tablehead{
  \colhead{Parameter}  & \colhead{2004}  & \colhead{2005}  & \colhead{2008}
}
\startdata
% updated 2011-11-05
      $c_{0}$  &     +0.00298  $\pm$ 0.00080  &     +0.00048  $\pm$ 0.00078  &     +0.00278  $\pm$ 0.00073  \\
      $c_{1}$  &     +0.00793  $\pm$ 0.00078  &     +0.01103  $\pm$ 0.00080  &     +0.01331  $\pm$ 0.00071  \\
      $c_{2}$  &     +0.00619  $\pm$ 0.00079  &     +0.00349  $\pm$ 0.00078  &     +0.00397  $\pm$ 0.00074  \\
      $c_{3}$  &     +0.00943  $\pm$ 0.00075  &     +0.00879  $\pm$ 0.00078  &     +0.00887  $\pm$ 0.00071  \\
      $c_{4}$  &     +0.00048  $\pm$ 0.00108  &     +0.00186  $\pm$ 0.00102  &     -0.00002  $\pm$ 0.00074  \\
      $c_{5}$  &     +0.01148  $\pm$ 0.00080  &     +0.01151  $\pm$ 0.00079  &     +0.01480  $\pm$ 0.00069  \\
      $c_{6}$  &     -0.00557  $\pm$ 0.00079  &     -0.00461  $\pm$ 0.00084  &     -0.00432  $\pm$ 0.00072  \\
      $c_{7}$  &     -0.00571  $\pm$ 0.00082  &     -0.00582  $\pm$ 0.00079  &     -0.00514  $\pm$ 0.00072  \\
      $c_{8}$  &     -0.00171  $\pm$ 0.00078  &     -0.00020  $\pm$ 0.00082  &     -0.00329  $\pm$ 0.00070  \\
      $c_{9}$  &     -0.00231  $\pm$ 0.00079  &     -0.00361  $\pm$ 0.00079  &     -0.00754  $\pm$ 0.00070  \\
     $c_{10}$  &     +0.00052  $\pm$ 0.00082  &     +0.00084  $\pm$ 0.00093  &     -0.00137  $\pm$ 0.00080  \\
     $c_{11}$  &     -0.00691  $\pm$ 0.00081  &     -0.00973  $\pm$ 0.00076  &     -0.00712  $\pm$ 0.00073  \\
     $c_{12}$  &     -0.00507  $\pm$ 0.00082  &     -0.00402  $\pm$ 0.00078  &     -0.00333  $\pm$ 0.00072  \\
     $c_{13}$  &     -0.01142  $\pm$ 0.00079  &     -0.00968  $\pm$ 0.00081  &     -0.01121  $\pm$ 0.00071  \\
  $F_*$ [mJy]  &    +18.78683  $\pm$ 0.00486  &    +18.70529  $\pm$ 0.00485  &    +18.60906  $\pm$ 0.00472  \\
\multicolumn{2}{r}{    $F_p/F_*$}  &  \multicolumn{2}{l}{     +0.00338  $\pm$ 0.00026}  \\
\multicolumn{2}{r}{    $T_{B}$ [K]}  &  \multicolumn{2}{l}{     +1310  $\pm$ 80}  \\
\multicolumn{2}{r}{$F_p$ [$\mu$Jy]}  &  \multicolumn{2}{l}{    +63.2  $\pm$ 4.9}  \\
\multicolumn{2}{r}{$T_{c,e}$ [BJD $_{{TDB}} $]\tablenotemark{b}}  &  \multicolumn{2}{l}{2453706.0595  $\pm$ 0.0014}  \\
\enddata
\tablenotetext{a}{Computed from the posterior MCMC distributions of
  $F_*$ and $F_p/F_*$.}  
\tablenotetext{b}{Jointly-fit ephemeris, assuming a period of 3.5247455~d
  \citep{torres:2008}.}
\end{deluxetable}

\begin{deluxetable}{l c c c}
\tablewidth{0pt}
\tabletypesize{\scriptsize}
\tablecaption{Independent Eclipse Fits \label{tab:indepeclipse}}
\tablehead{
  \colhead{Parameter}  & \colhead{2004}  & \colhead{2005}  & \colhead{2008}
}
\startdata
% Updated 2011-11-05
$c_{0}$  &  0.00262  $\pm$ 0.00090       &  -0.00020  $\pm$ 0.00112   &  0.00044  $\pm$ 0.00264   \\
$c_{1}$  &  0.00744  $\pm$ 0.00089       &  0.01028  $\pm$ 0.00124    &  0.01246  $\pm$ 0.00114   \\
$c_{2}$  &  0.00500  $\pm$ 0.00169       &  0.00396  $\pm$ 0.00084    &  0.00449  $\pm$ 0.00106   \\
$c_{3}$  &  0.00879  $\pm$ 0.00101       &  0.00879  $\pm$ 0.00077    &  0.00850  $\pm$ 0.00075   \\
$c_{4}$  &  -0.00042  $\pm$ 0.00090      &  0.00198  $\pm$ 0.00111    &  0.00143  $\pm$ 0.00160   \\
$c_{5}$  &  0.01090  $\pm$ 0.00104       &  0.01109  $\pm$ 0.00091    &  0.01392  $\pm$ 0.00115   \\
$c_{6}$  &  -0.00488  $\pm$ 0.00106      &  -0.00502  $\pm$ 0.00109   &  -0.00351  $\pm$ 0.00120  \\
$c_{7}$  &  -0.00453  $\pm$ 0.00138      &  -0.00331  $\pm$ 0.00284   &  -0.00561  $\pm$ 0.00086  \\
$c_{8}$  &  -0.00248  $\pm$ 0.00109      &  -0.00078  $\pm$ 0.00089   &  -0.00249  $\pm$ 0.00105  \\
$c_{9}$  &  -0.00230  $\pm$ 0.00082      &  -0.00239  $\pm$ 0.00142   &  -0.00744  $\pm$ 0.00073  \\
$c_{10}$  &  0.00113  $\pm$ 0.00118       &  0.00086  $\pm$ 0.00093    &  0.00079  $\pm$ 0.00225  \\
$c_{11}$  &  -0.00588  $\pm$ 0.00118      &  -0.01021  $\pm$ 0.00097   &  -0.01012  $\pm$ 0.00328 \\
$c_{12}$  &  -0.00315  $\pm$ 0.00201      &  -0.00385  $\pm$ 0.00078   &  -0.00189  $\pm$ 0.00171 \\
$c_{13}$  &  -0.01197  $\pm$ 0.00104     &  -0.01090  $\pm$ 0.00166    &  -0.01059  $\pm$ 0.00098 \\
$T_{c,e}$  [BJD $_{{TDB}} $]\tablenotemark{a}&  2453346.5348  $\pm$ 0.0028  &  2453706.0600  $\pm$ 0.0029   &  2454675.3639  $\pm$ 0.0026\\
$F_p/F_*$  &  0.00325  $\pm$ 0.00053     &  0.00384  $\pm$ 0.00046    &  0.00281  $\pm$ 0.00051   \\
$T_{B}$ [K] & 1270  $\pm$ 190 & 1450$\pm$ 230 &  1130 $\pm$ 160 \\
$F_*$ [mJy]  &  18.7884  $\pm$ 0.0069  &  18.7023  $\pm$ 0.0064   &  18.6155  $\pm$ 0.0068  \\
$F_p$  \tablenotemark{b} [$\mu$Jy]&  61.1  $\pm$ 10.0        &  71.9  $\pm$ 8.6   &  52.3  $\pm$ 9.5  \\

\enddata
\tablenotetext{a}{Jointly-fit ephemeris, assuming a period of 3.5247455~d
  \citep{torres:2008}.}

\tablenotetext{b}{Computed from the posterior MCMC distributions of
  $F_*$ and $F_p/F_*$.}
\end{deluxetable}

\clearpage

\end{document}